\documentclass[a4paper,fleqn]{cas-sc}

\usepackage{epsfig}
\usepackage[numbers]{natbib}
\usepackage{amssymb}
\usepackage{amsmath}
\usepackage{amsthm}

\usepackage{rotating}
\usepackage[T1]{fontenc}
\usepackage{color}
\usepackage{xcolor}
\usepackage{multirow}
\usepackage{afterpage}
\usepackage{todonotes}
\usepackage{soul}
\usepackage{textcomp}

\def\tsc#1{\csdef{#1}{\textsc{\lowercase{#1}}\xspace}}
\tsc{WGM}
\tsc{QE}

\newcommand{\ok}{\textcolor{green}{\textbf{\checkmark}}}
\newcommand{\conditional}{\textcolor{orange}{\textbf{!}}}
\newcommand{\notok}{\textcolor{red}{\textbf{\texttimes}}}

\begin{document}
\let\WriteBookmarks\relax
\def\floatpagepagefraction{1}
\def\textpagefraction{.001}

\shorttitle{Developing Guidelines for XAI Evaluation}    

\shortauthors{Velmurugan et al}  

\title [mode = title]{Developing Guidelines for Functionally-Grounded Evaluation of Explainable Artificial Intelligence using Tabular Data}  



%

\author[inst1]{Mythreyi Velmurugan}[orcid=0000-0002-5017-5285]
\credit{Conceptualisation, Methodology, Investigation, Writing - Original Draft, Writing - Review \& Editing}

\author[inst1]{Chun Ouyang}[orcid=0000-0001-7098-5480]
\credit{Conceptualisation, Methodology, Investigation, Writing - Review \& Editing, Supervision}

\author[inst1]{Yue Xu}[orcid=0000-0002-1137-0272]
\credit{Investigation, Writing - Review \& Editing}

\author[inst2]{Renuka Sindhgatta}[orcid=0000-0001-7533-533X]
\credit{ Conceptualisation, Methodology, Investigation}

\author[inst1]{Bemali Wickramanayake}[orcid=0000-0001-6480-3513]
\credit{Investigation}

\author[inst1,inst3]{Catarina Moreira}[orcid=0000-0002-8826-5163]
\credit{Conceptualisation, Methodology, Investigation}

\affiliation[inst1]{organization={Queensland University of Technology},
            addressline={2 George St}, 
            city={Brisbane},
            postcode={4000}, 
            state={Queensland},
            country={Australia}}

\affiliation[inst2]{organization={IBM Research},
            addressline={Manyata Embassy Business Park, Outer Ring Road}, 
            city={Bengaluru},
            postcode={560045}, 
            state={Karnataka},
            country={India}}

\affiliation[inst3]{organization={University of Technology Sydney},
            addressline={15 Broadway}, 
            city={Ultimo},
            postcode={2007}, 
            state={New South Wales},
            country={Australia}}

\cortext[cor1]{Corresponding Author: m.velmurugan@qut.edu.au}

















\begin{abstract}
Explainable Artificial Intelligence (XAI) techniques are used to provide transparency to complex, opaque predictive models. However, these techniques are often designed for image and text data, and it is unclear how fit-for-purpose they are when applied to tabular data. As XAI techniques are rarely evaluated in settings with tabular data, the applicability of existing evaluation criteria and methods are also unclear and needs (re-)examination. For example, some works suggest that evaluation methods may unduly influence the evaluation results when using tabular data. This lack of clarity on evaluation procedures can lead to reduced transparency and ineffective use of XAI techniques in real world settings. In this study, we examine literature on XAI evaluation to derive guidelines on functionally-grounded assessment of local, post hoc XAI techniques. 
We identify 20 evaluation criteria and associated evaluation methods, and derive guidelines on when and how each criterion should be evaluated. We also identify key research gaps to be addressed by future work. 
Our study contributes to the body of knowledge on XAI evaluation through in-depth examination of functionally-grounded XAI evaluation protocols, and has laid the groundwork for future research on XAI evaluation. 
\end{abstract}


\begin{highlights}
\item Investigation of functionally-grounded XAI evaluation when using tabular data
\item Twenty evaluation criteria and six dimensions of quality are identified
\item Guidelines on conducting functionally-grounded evaluation are extracted
\item Emerging areas of research and key future directions are identified
\end{highlights}


\maketitle

\section{Introduction}\label{sec:intro}
Modern machine learning relies on sophisticated algorithms to produce accurate predictive models. However, this sophistication results in complex models whose workings are opaque to users. Post hoc explainable AI (XAI) techniques are applied to provide transparency to these models~\cite{Veran2023}. However, the goodness of these techniques is still being debated~\cite{Rudin2019}. This is particularly true when applied to models trained on tabular data, as most XAI techniques are designed for image and text, then adapted to tabular data~\cite{DiMartino2022}.

Tabular data have unique characteristics that make prediction and explanation challenging. Tabular datasets are highly heterogeneous with respect to the quantity and type of features they may have, and may present challenges such as sparsity in data, lack of locality and the need to manage mixed feature types~\cite{ShwartzZiv2022}. Furthermore, there may be fewer correlations between features in tabular data, as well as lack of semantic meaning and connection between features~\cite{Borisov2021}. Despite the unique characteristics of tabular data, XAI techniques applied to tabular data, and tabular time series data in particular, are typically adapted from those used in image processing or natural language processing~\cite{DiMartino2022}. It has been noted that one of the key challenges for XAI techniques is the accuracy and relevance of explanations, especially when using tabular data~\cite{Sahakyan2021}. 

Nevertheless, explanations are comparatively rarely assessed for these functional qualities in the literature using tabular data. Dieber and Kirrane~\cite{Dieber2022} find that even LIME, one of the most commonly applied post hoc XAI techniques, is typically used and assessed on image data. 
It is unclear whether existing XAI evaluation methods and metrics can be applied when using tabular data. 
Existing literature focusing on tabular data often adapt evaluation methods used to assess explanations for models trained on image and text data. 
Whilst several works have assessed explanation fidelity (i.e., how faithful an explanation is to the model) by adapting an existing evaluation method~\cite{Naretto2022,Ozyegen2021,Veerappa2022}, it has been revealed that the adapted evaluation method is highly susceptible to the specific parameters of the evaluation procedure, and that results often change when evaluation parameters are adjusted~\cite{Maltbie2021,Veerappa2022}. These findings suggest that any results produced from assessment may be attributable to the evaluation method, rather than the explanation quality, and 
thus highlights the need to ensure that evaluation methods and metrics are appropriate for a given technical context. 
%

To this end, we identify a significant gap in the literature concerning the evaluation of explanations for tabular data. 
The problem is three-fold. 
Firstly, it is unclear \textit{what} to evaluate due to the lack of consensus on what constitutes a good explanation, i.e., the criteria that explanations should meet. 
Secondly, it is unclear \textit{how} to evaluate due to the lack of standardised methods and metrics to assess the quality and effectiveness of explanations. Current practices are fragmented, with researchers and practitioners using diverse and often incomparable evaluation techniques. 
Thirdly, it is unclear \textit{when} to evaluate due to the fact that the context of evaluation is often overlooked. 
The appropriateness of certain criteria and evaluation methods can vary depending on the specific evaluation purpose, data modality, and inherent characteristics of machine learning models and XAI techniques. 
The lack of clarity on suitable evaluation procedures not only hampers the ability to benchmark and improve XAI methods effectively, but also can lead to reduced transparency and ineffective use of XAI techniques in real-world applications. Consider the following two scenarios.  



In the first scenario, University A is developing a machine learning solution to identify at-risk and disadvantaged students and provide necessary interventions. Student success may be influenced by various factors, such as socio-economic status, psychological pressure, and prior educational experiences~\cite{Dart2019}. Therefore, it is crucial for the university to direct students to the most appropriate mentorship or support programs to provide personalised and timely assistance~\cite{Khan2021}, and it is essential for employees to understand why a student is flagged by a predictive algorithm. To address this need, University A plans to apply a local, post hoc XAI technique to the predictive algorithm to identify the features that led to a student being flagged. However, given the plethora of XAI techniques available, the university aims to determine how to shortlist the most suitable XAI techniques for further assessment. As such, they require clarity on the criteria that can be used to evaluate the suitability of an XAI technique for their specific objectives. 

In the second scenario, Company B is using predictive process analytics (PPA) to monitor business processes and identify potential undesirable future states. PPA involves complex tabular data with dependencies between features and temporal dependencies between records, often resulting in opaque models~\cite{Velmurugan2021a,Wickramanayake2023}. To enhance transparency, Company B plans to apply post hoc XAI techniques to interpret their models. However, previous evaluations of XAI techniques applied to PPA have suggested that these techniques may not be fully fit-for-purpose~\cite{Stevens2024,Velmurugan2021a}. As a result, Company B intends to assess the faithfulness of several XAI techniques with respect to their process predictive model, focusing on: (1) the completeness of feature attribution explanations (i.e., whether other features are needed to produce the same prediction), and (2) the necessity of specific features for a prediction (i.e., whether the same prediction can be made without a given feature). Given the variety of evaluation methods and metrics for XAI techniques, Company B requires guidance on assessing these properties effectively.

%

Based on the above, addressing the identified gap in the evaluation of explanations for tabular data is imperative for XAI. Establishing clear criteria, methods, and contextual guidelines will significantly enhance the transparency, reliability, and utility of XAI techniques, ultimately leading to their more effective and widespread deployment. 
Therefore, in this paper, we investigate the evaluation of post hoc XAI techniques in settings that use tabular data. More specifically, we investigate how functionally-grounded evaluation can be conducted, 
and this form of evaluation assesses the inherent goodness of XAI techniques without users. As the evaluation methods and metrics that may be used to assess explanations appear to be highly specific to the technical context, we take a targeted approach and analyse explanation evaluation methods in depth. We develop guidelines for evaluating explanations, with a focus on functionally-grounded evaluation and on models trained using tabular data. Our guidelines fulfill the following objectives: 
\begin{itemize}
    \item identify functionally-grounded criteria that explanations must meet;
    \item identify evaluation methods and metrics that have been used to assess explanations against the identified criteria; and 
    \item identify evaluation methods and their appropriateness for a technical context for each criterion.
\end{itemize}
In addition to the above, our guidelines also specify opportunities for further investigation and potential improvements to evaluation methods that can be addressed by future work.

We begin with an overview of key surveys and systematic literature reviews that describe XAI evaluation (Section~\ref{sec:background}), then explain the approach used in this work (Section~\ref{sec:Method}). We present our taxonomy of evaluation criteria and evaluation methods (Section~\ref{sec:taxonomy}), analyse each criterion in detail (Section~\ref{sec:fidelity-slr} to Section~\ref{sec:utility-slr}), and discuss our contributions and further research gaps 
(Section~\ref{sec:discussion}). Finally, we conclude the paper (Section~\ref{sec:conclusion}).

\section{Related Works}\label{sec:background}

While several existing studies conduct systematic literature reviews on XAI, few examine explanation evaluation in-depth. For example, some key works place emphasis on identifying and categorising XAI techniques, and present no or only a small overview of XAI evaluation~\cite{DiMartino2022,Guidotti2018,Owens2022,Theissler2022}. Additionally, many works merely categorise evaluation criteria -- the conditions that explanations must meet -- and do not examine evaluation procedures -- namely, the methods and metrics. For example, one seminal work in XAI evaluation~\cite{DoshiVelez2017} presents a three-level taxonomy for evaluation criteria:
\begin{itemize}
    \item \textit{Functionally-grounded evaluation}, in which the inherent qualities of an explanation are assessed with no subjective, user-based evaluation;
    \item \textit{Human-grounded evaluation}, in which an explanation is assessed in a proxy setting, with laypeople doing simple or simplified tasks; and
    \item \textit{Application-grounded evaluation}, in which an explanation is assessed in full context, with end users.
\end{itemize}
However, this taxonomy does not comprehensively include the criteria that may fall under each of these levels, or how these criteria are evaluated.


Some works categorise evaluation criteria and evaluation methods without consideration of a specific context. These works have some notable exclusions and gaps. Firstly, while there is some overlap between the taxonomies produced by these works, there are also notable inconsistencies. For example, Carvalho, Pereira and Cardoso~\cite{Carvalho2019} define explanation completeness as related to the number of data points covered by an explanation, but Zhou et al~\cite{Zhou2021} define completeness as a measure of model behaviour coverage. Similarly, while Mohseni, Zarei and Ragan~\cite{Mohseni2021} refer to explanation fidelity as a single criterion, their definition aligns more with the completeness criterion proposed by Zhou et al~\cite{Zhou2021}, which is described as only one aspect of fidelity. Furthermore, there are a number of criteria or categories of criteria that are present in only some works, but not others. Additionally, most existing works focus only on \textit{what} evaluation criteria exist should be used to assess explanations. Few works consider in detail \textit{how} these criteria should be evaluated. 
To the best of our knowledge, existing works consider XAI evaluation more generally and often in a broader context, whereas no work makes a targeted and in-depth survey of XAI evaluation for tabular data that can be adapted for operationalisation.

\begin{table}[]
    \centering
    \caption{Existing taxonomies in literature often have inconsistencies that must be addressed when building upon them. In particular, some criteria are given different names across different taxonomies, and criteria with the same name may be given different definitions across taxonomies. We summarise some key taxonomies, including overlaps in name and/or defintion, below.}
    \label{tab:background-summary}
    \begin{tabular}{|p{0.2\linewidth}|p{0.6\linewidth}|p{0.1\linewidth}|}
    \hline
    \textbf{Criterion Name} &
      \textbf{Criterion Definition} &
      \textbf{Citation} \\ \hline
    Sensitivity &
      Features with a non-zero impact on the model should have a non-zero attribution and vice versa &
      \cite{Carvalho2019} \\ \hline
    Soundness &
      How correct and truthful an explanation is. \cite{Zhou2021}~\cite{Zhou2021} describes it as one aspect of fidelity. &
      \cite{Zhou2021} \\ \hline
    \multirow{2}{\linewidth}{Correctness} &
      \multirow{4}{\linewidth}{Accuracy   of explanation in how it describes the model. \cite{Zhou2021}~\cite{Zhou2021} describes it a group of criteria rather than a singular criterion.} &
      \cite{Carvalho2019} \\ \cline{3-3} 
     &
       &
      \cite{Nauta2023} \\ \cline{1-1} \cline{3-3} 
    \multirow{2}{\linewidth}{Fidelity} &
       &
      \cite{Zhou2021} \\ \cline{3-3} 
     &
       &
      \cite{Mohseni2021} \\ \hline
    Implementation\hspace{1em} Invariance &
      \multirow{2}  {\linewidth}{Two models trained on the same data and making the same predictions should have identical explanations} &
      \cite{Carvalho2019} \\ \cline{1-1} \cline{3-3} 
    Consistency &
       &
      \cite{Nauta2023} \\ \hline
    Separability &
      Non-identical data points should have non-identical explanations &
      \cite{Carvalho2019} \\ \hline
    Stability &
      \multirow{2}  {\linewidth}{Similar data points should have similar explanations} &
      \cite{Carvalho2019} \\ \cline{1-1} \cline{3-3} 
    Continuity &
       &
      \cite{Nauta2023} \\ \hline
    Robustness &
      Explanation is stable when the input is slightly perturbed &
      \cite{Clement2023} \\ \hline
    Sanity Check &
      Measures how   strongly an explanation reacts to randomisation in model parameters or data &
      \cite{Clement2023} \\ \hline
    \multirow{3}  {\linewidth}{Completeness} &
      \multirow{4}  {\linewidth}{Coverage   of explanations/amount of model behaviour described by an explanation.   Additionally, \cite{Zhou2021} identifies it as a facet of interpretability (as broadness)  and fidelity (as completeness).} &
      \cite{Nauta2023} \\ \cline{3-3} 
     &
       &
      \cite{Carvalho2019} \\ \cline{3-3} 
     &
       &
      \cite{Zhou2021} \\ \cline{1-1} \cline{3-3} 
    Broadness &
       &
      \cite{Zhou2021} \\ \hline
    Complexity &
      \multirow{2}  {\linewidth}{Size and complexity of explanation} &
      \cite{Clement2023} \\ \cline{1-1} \cline{3-3} 
    \multirow{2}  {\linewidth}{Compactness} &
       &
      \cite{Nauta2023} \\ \cline{2-3} 
     &
      \multirow{2}  {\linewidth}{Requires that explanations be succinct. Additionally, Zhou et al~\cite{Zhou2021} identifies it as a facet of interpretability.} &
      \cite{Carvalho2019} \\ \cline{1-1} \cline{3-3} 
    Parsimony &
       &
      \cite{Zhou2021} \\ \hline
    Interpretability &
      A group of   criteria which measure how understandable an explanation is &
      \cite{Zhou2021} \\ \hline
    Clarity &
      Ambiguousness of   an explanation &
      \cite{Zhou2021} \\ \hline
    Task-specific evaluation &
      \multirow{2}  {\linewidth}{Usefulness   of explanation in the context of the specific purpose in which it will be   used} &
      \cite{Clement2023} \\ \cline{1-1} \cline{3-3} 
    Context &
       &
      \cite{Nauta2023} \\ \hline
    Method-specific\hspace{2em} evaluation &
      Measures an   inherent property specific to a particular explanation or type of explanation &
      \cite{Clement2023} \\ \hline
    Model Trustworthiness &
      How easy is it   to determine the model's trustworthiness using explanations &
      \cite{Mohseni2021} \\ \hline
    Contrastivity &
      How well an   explanation describes why a different explanation was not returned &
      \cite{Nauta2023} \\ \hline
    Covariate Complexity &
      How complex   interactions of features are in the explanation, and how human-understandable   the concepts are &
      \cite{Nauta2023} \\ \hline
    Composition &
      How well an   explanation is organised and presented &
      \cite{Nauta2023} \\ \hline
    Confidence &
      Whether   probability information is presented in the explanation, and if so, how   accurate this information is. &
      \cite{Nauta2023} \\ \hline
    Coherence &
      \multirow{2}  {\linewidth}{How much an explanation concurs with prior   knowledge and beliefs, and real-world importance of features} &
      \cite{Nauta2023} \\ \cline{1-1} \cline{3-3} 
    Faithfulness &
       &
      \cite{Clement2023} \\ \hline
    Controllability &
      How interactive   an explanation is &
      \cite{Nauta2023} \\ \hline
    \end{tabular}
\end{table}

To address the above gaps and inconsistencies, we first synthesise a taxonomy that 1) standardises names and definitions for evaluation criteria and 2) examines the evaluation methods that may be applied to each criterion. While we refer to existing taxonomies when creating our own, we also need to bridge the gaps we identify in existing taxonomies, standardise terminology and include the state-of-the-art in explanation evaluation. Secondly, we develop guidelines for XAI evaluation. For each criterion in our taxonomy, we extract relevant details of evaluation from existing works, and develop guidelines on when and how each criterion should be evaluated. More specifically, the overarching research problem that we aim to address is: \textit{How can explanations of predictive models for tabular data (including time series data) be evaluated and compared?} 
As a starting point, we examine below and in Table~\ref{tab:background-summary} several key systematic literature reviews and surveys on XAI evaluation.

Similar to Doshi-Velez and Kim~\mbox{\cite{DoshiVelez2017}}, Carvalho, Pereira and Cardoso~\mbox{\cite{Carvalho2019}} categorise XAI evaluation criteria into \textit{qualitative interpretability indicators} and \textit{quantitative interpretability indicators}. 
{This early work identifies eight criteria that are commonly evaluated in the XAI literature. However, as newer evaluation criteria and evaluation taxonomies have since emerged, the taxonomy proposed by Carvalho, Pereira and Cardoso~\mbox{\cite{Carvalho2019}} contains omissions and gaps when compared to newer taxonomies.}

{Clement et al~\mbox{\cite{Clement2023}} classify evaluation criteria into \textit{computational} and \textit{human-based} evaluation. This work identifies six specific evaluation criteria.}
{While Clement et al~\mbox{\cite{Clement2023}} provide an overview of methods and metrics that may be used to evaluate explanations, they do not offer guidance on which criteria or procedures can be applied under which circumstances. The criteria themselves are also somewhat inconsistent. For example, while robustness and faithfulness are very specific and define the properties they measure, task-specific and method-specific evaluation are less clear.}


{Zhou et al~\mbox{\cite{Zhou2021}} identify six functionally-grounded criteria sorted in two groups:}
\begin{itemize}
    \item {Interpretability: criteria that measure how understandable an explanation is~\mbox{\cite{Zhou2021}}.} 
    \item {Fidelity: criteria that measure how accurately explanations describe the model~\mbox{\cite{Zhou2021}}.} 
\end{itemize}
{Zhou et al~\mbox{\cite{Zhou2021}} identify evaluation procedures for each criterion, and the types of explanations that they may be applied to. However, the effectiveness and limitations of the criteria and the evaluation methods with respect to different types of data and models are not discussed.}

{Mohseni, Zarei and Ragan~\mbox{\cite{Mohseni2021}} examine both the evaluation criteria and evaluation methods, emphasising human-grounded and application-grounded evaluation. This work identifies only two functionally-grounded criteria.}
{Moreover, while Mohseni, Zarei and Ragan~\mbox{\cite{Mohseni2021}} provide a brief summary of evaluation methods applied in literature, they do not examine these methods in detail, nor do they offer guidelines on how to carry out these evaluation methods.}

{Most recently, Nauta et al~\mbox{\cite{Nauta2023}} identify twelve explanation criteria that explanations must meet, each of which is related to the content of the explanation, the presentation of the explanation or the user of the explanation.}
{While} Nauta et al~\mbox{\cite{Nauta2023}} identify evaluation methods to assess each criterion, their work addresses a broader context. They do not distinguish between evaluation procedures used for different experimental settings, how the machine learning pipeline may affect explanation evaluation or how evaluation may be operationalised.

\section{Method}\label{sec:Method}

\begin{figure}[tb!!!]
    \centering
    \includegraphics[scale=0.7]{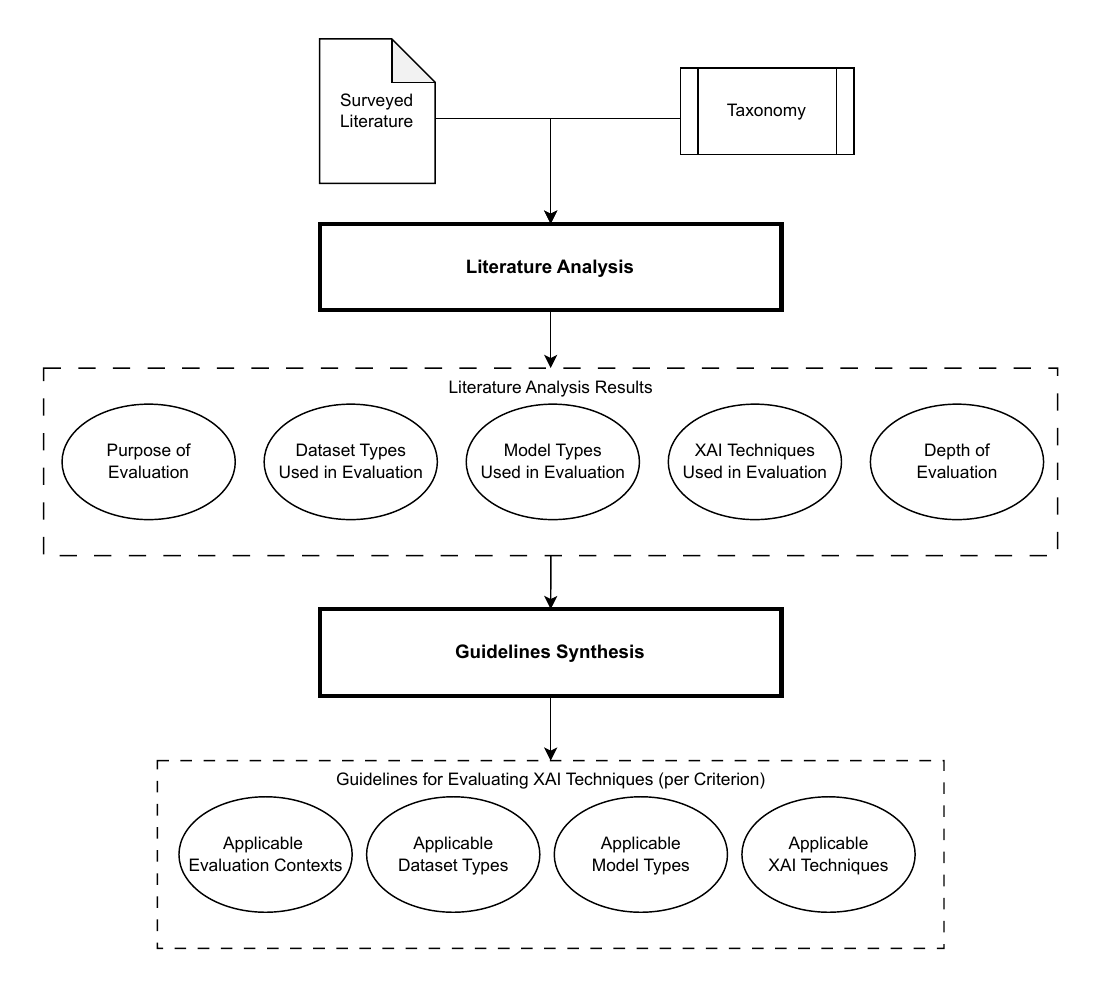}
    \caption{An overview of our multi-phase approach, starting with a systematic literature review to synthesise guidelines for XAI evaluation.}
    \label{fig:approach}
\end{figure}

We conduct our work in several phases (see Figure~\ref{fig:approach}). We begin with a systematic literature review to identify evaluation criteria, methods and metrics used in existing literature. We then analyse the surveyed works in depth to determine how and when these evaluation procedures are applied, and whether any limitations to application of these procedures exist. We then use this analysis to synthesise guidelines for XAI evaluation.

\begin{figure}[tb!!!]
    \centering
    \includegraphics[scale=0.7]{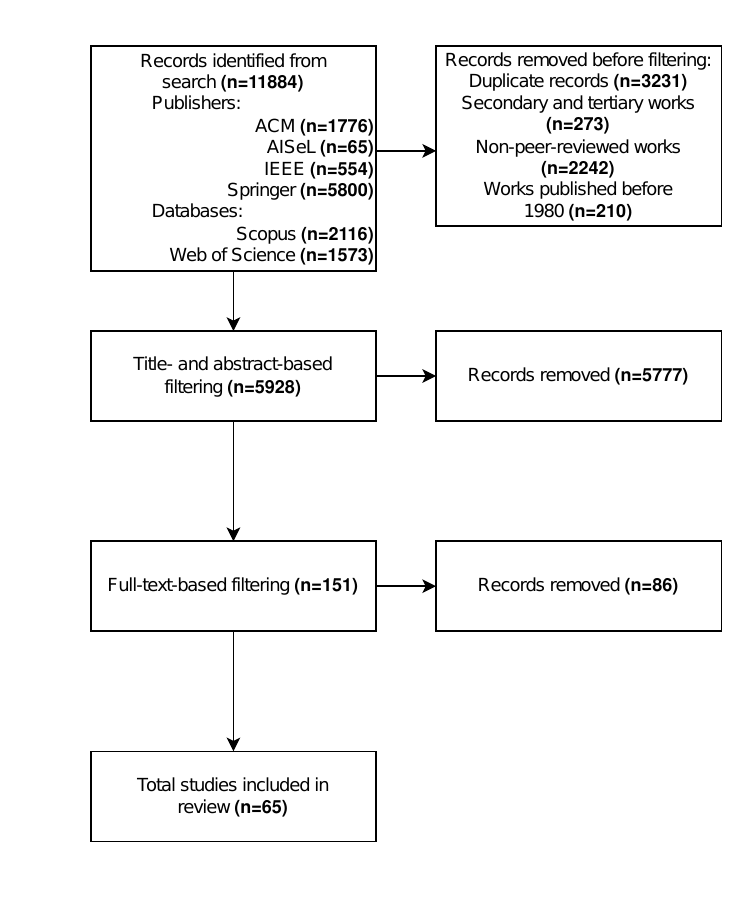}
    \vspace*{-1.5\baselineskip}
    \caption{The search and filtering of literature for the systematic review guided by the PRISMA methodology.}
    \label{fig:prisma}
\end{figure}

\textbf{Literature Search and Selection.}
The PRISMA methodology~\cite{Page2021} was applied when conducting the systematic literature review (see Figure~\ref{fig:prisma}). Two research publication databases and four publishers' databases were searched using the following keywords: (``explainable ai''  OR  xai  OR ``interpretable ai''  OR ``interpretable machine learning'' OR ``explainable machine learning'') AND (quality OR evaluat* OR metrics). 

Papers published between 1 January 1980 and 6 February 2023 were collected via the search, resulting in 11884 results. Firstly, pre-filtering was conducted to remove duplicates and non-peer-reviewed results. 
We found that a vast number of papers were indexed across multiple databases and publishers. 
Also, some searches had returned many non-peer-reviewed sources. After this pre-filtering, 5928 papers remained. 

We used several inclusion and exclusion criteria to filter the remaining papers based on relevance to our research aims. 
{Papers which met all of the following inclusion criteria were considered in the systematic review}:
\begin{enumerate}
    \item Papers that contain primary studies, including bibliographic studies, systematic literature reviews and surveys that are followed by primary studies.
    \item Papers that investigate assessment of explanations for supervised learning tasks.
    \item Papers that investigate assessment  local, post hoc explainable methods on tabular data.
    \item {Papers that conduct functionally-grounded evaluation of XAI techniques, or that describe detailed evaluation methods and metrics without conducting evaluation}.
\end{enumerate}

{Papers that met any of the following criteria were excluded}:
\begin{enumerate}
    \item Papers that evaluate predictive models using explanations, but do not evaluate explanations.
    \item Papers that describe explanation evaluation criteria or requirements that explanations must meet, but not evaluation methods or metrics.
    \item Papers that only conduct non-functional evaluations or sanity checks.
    \item Papers evaluating only algorithms that produce only counterfactual or contrastive explanations (these forms of explanations have specific requirements that must be met, and so are typically evaluated using distinct criteria~\cite{Chou2021fusion}).
\end{enumerate}

{In our inclusion criteria, we specify that we include only works evaluating post hoc approaches, and exclude inherently interpretable, ``transparent'' models. While such models are a key form of providing explainability~\mbox{\cite{Guidotti2018}}, we focus on post hoc XAI, as it is typically the evaluation of post hoc explanations which have been adapted from other forms of data. Moreover, many of the criteria applied to post hoc explanations are not applicable to transparent models, such as an explanation's fidelity to the model.}

Two rounds of paper filtering were conducted during this phase. 
Firstly, all papers were filtered using {both} the title and the abstract, reducing the number of papers to 151. {Many works were excluded at this stage, which were not relevant to XAI (or even the field of information technology). For example, several of the excluded works were geological studies set in Xai-Xai in Mozambique).} 
Secondly, the remaining papers were filtered again for relevance based on the full text. This resulted in the final identification of 65 papers. 


\textbf{Identification of Evaluation Procedures.}
Each paper is analysed with respect to the \textit{criteria} used to assess (quality of) explanations and explanation \textit{evaluation methods}. We group together criteria that share the same goal when evaluating, which we refer to as \textit{dimensions of explanation quality}. Using these dimensions, criteria and evaluation methods, we build a taxonomy of XAI evaluation. 
We only consider criteria that are not specific to a single XAI method. We exclude, for example, the evaluation of consistency property that is inherent only to SHAP explanations~\cite{Lundberg2020}. 

\textbf{Synthesis of Guidelines.}
{For each criterion, the surveyed literature was analysed with respect to the following:}
\begin{itemize}
    \item {The context and purpose of the evaluation: When analysing the surveyed literature, it is apparent that the context of the evaluation falls into two, broad categories. Most works conduct \textit{general-purpose} evaluation without consideration of a specific domain or task (e.g. a benchmark study or comparison of new and existing XAI techniques). Some works conduct evaluation in a setting specific to a prediction task or domain, i.e. some form of \textit{application-based evaluation}.}
    \item {The type(s) of tabular data used: We identify four types of tabular data used in the surveyed literature. The four types of data we identify are real-life \textit{tabular}, \textit{univariate time series}, \textit{multivariate time series} and \textit{synthetic tabular}. We include time series data in our work as this form of data often has some similarity to tabular data, such as sparsity in the input, lack of locality and the need to manage mixed feature types.}
    \item {The type(s) of predictive models used: We identified numerous, broad categories of predictive models that are used to evaluate explanations in literature. We examine the effect of these predictive models on the evaluation criteria and evaluation methods, and where necessary we specify these categories.}
    \item {The type(s) of local explanations or XAI techniques: We identify four categories of explanations tested in the surveyed literature. The four categories we identify are \textit{feature attribution} explanations, \textit{rule-based} explanations, \textit{feature profiling} explanations (for example, Partial Dependence Plot (PDP)) and \textit{feature sets}. We do not consider explanations derived from transparent models, and counterfactual explanations. Although both are important and common forms of providing explainability, they must meet different requirements from factual post hoc explanations. 
    Thus, we exclude these two branches of XAI techniques from our work.}
\end{itemize}

{We include our analysis in the supplementary materials. With the taxonomy we propose (Section~\mbox{\ref{sec:taxonomy}}), we analyse the surveyed literature and derive guidelines on assessing each of the identified criteria (Sections~\mbox{\ref{sec:fidelity-slr}} to~\mbox{\ref{sec:utility-slr}}).}


\section{Taxonomy of XAI Evaluation}\label{sec:taxonomy}
We develop a taxonomy of evaluation methods to assess post hoc explanations. 
Our taxonomy is composed of three levels. The top level captures six \emph{dimensions of explanation quality}, which we sort evaluation criteria into. The second level is composed of \emph{evaluation criteria}. The third level indicates \emph{evaluation methods} that may be used to evaluate these criteria.
Through the systematic review, we identify 20 evaluation criteria that explanations can be assessed against. These criteria can be grouped together according to a shared goal, which informs a dimension of explanation quality. For example, six criteria evaluate how well an explanation captures a model's behaviour (fidelity to the model), and three criteria evaluate how well an explanation captures reality (fidelity to the ground truth). As a result, we group these 20 criteria into six overarching dimensions of explanation quality. 

\begin{figure}[!htb]
    \centering
    \includegraphics[scale=0.45]{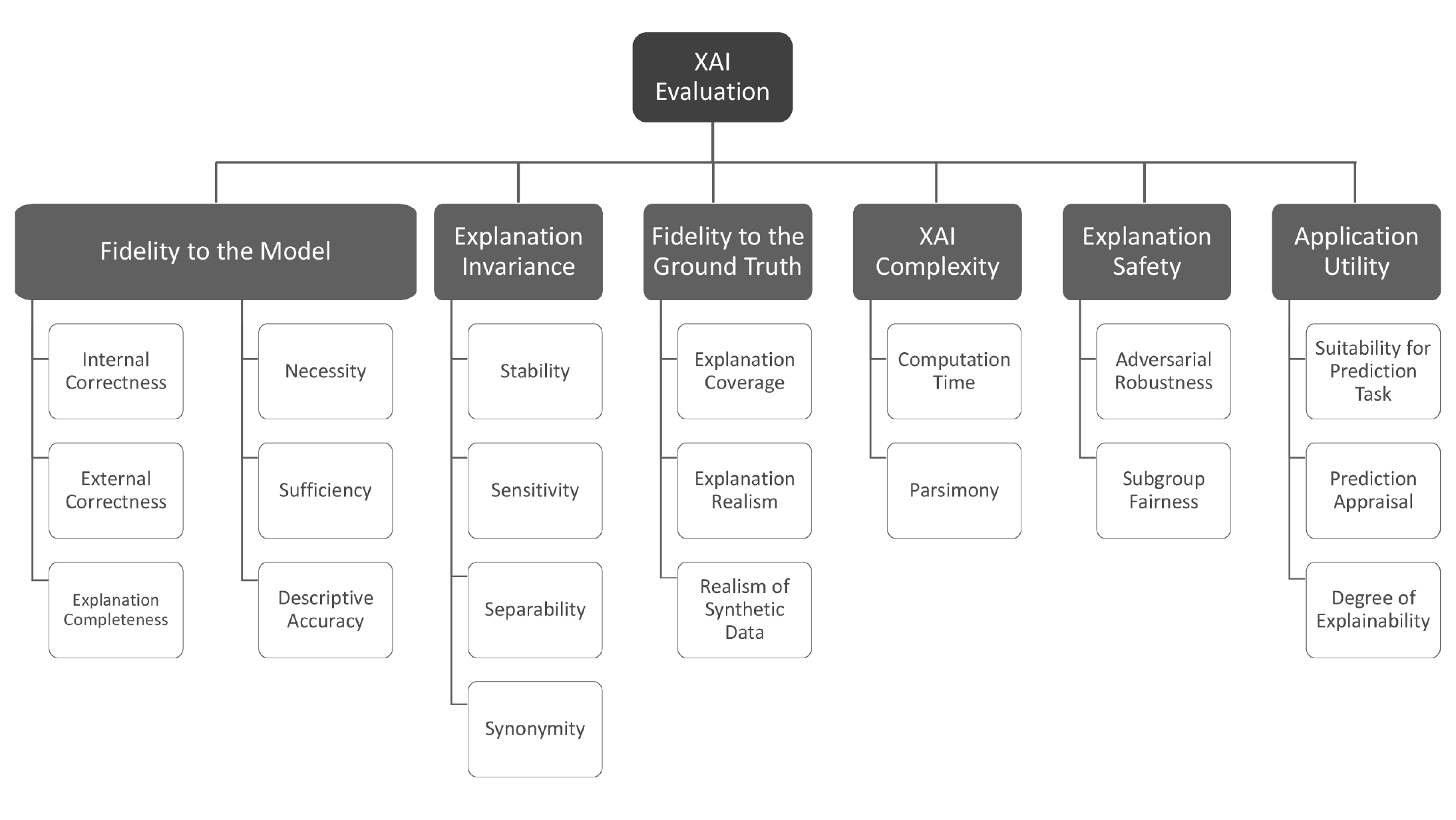}
    \caption{The first and second levels of the proposed taxonomy, showing the dimensions of explanation quality and the associated criteria. There are 20 criteria that may be used to assess explanations, which we group into six quality dimensions.}
    \label{fig:taxonomy_top_level}
\end{figure}

Figure~\ref{fig:taxonomy_top_level} depicts the first level (dimensions of explanation quality) and second level (evaluation criteria) of the proposed taxonomy. Below, we describe the six quality dimensions, respectively. 

\begin{itemize}
    \item \textbf{Fidelity to the model} is defined as how well an explanation approximates a black-box model~\cite{Carvalho2019,Mohseni2021}. Some works identify two aspects of fidelity: the completeness of explanations in describing the model; and the correctness of explanations~\cite{Sokol2020,Zhou2021}. Upon examining the works identified through our systematic literature review, we clarify the definitions of existing criteria and identify more criteria associated with fidelity to the model.
    
    \item \textbf{Explanation Invariance} is suggested by Sokol and Flach~\cite{Sokol2020} to be a necessary element for determining the effect of the model and/or explanation parameters on the explanation. Explanations should reflect how the model captures the phenomenon exemplified by the data, and should not be influenced by characteristics of the model or the explanation generation process. Thus, it is necessary to measure how explanations vary across a dataset and different predictive models. The name explanation invariance is applied to this dimension of explanation quality by Hailemariam et al~\cite{Hailemariam2020}.

    \item \textbf{Fidelity to Ground truth} is not proposed as a dimension in literature. Several of the surveyed works assess how well an explanation captures the training dataset, or how well it captures real life, rather than how well it reflects the model.  
    As the model does not necessarily correctly capture real-world relations, evaluating fidelity with respect to the ground truth does not guarantee explanation fidelity to the model. Thus, three criteria associated with fidelity are separated into this dimension. 

    \item \textbf{XAI Complexity} is not proposed in literature. We sort into this dimension two common measures of explanation quality that assess, respectively, the complexity of explanation generation and explanation complexity. 

    \item \textbf{Explanation Safety} is composed of criteria that are relatively new in the literature. An emerging area of research assesses the extent to which an explanation can be applied in practice without risk of harm to stakeholders. We identified two criteria that are used to perform this form of assessment. 

    \item \textbf{Application Utility} is briefly touched on by Clement et al~\cite{Clement2023} as task-specific evaluation. We expand this definition, and sort into this dimension any criteria that determine how effective an explanation is when applied in a  given domain or task context. 
\end{itemize}

Based on the above, we examine the individual evaluation criteria and relevant evaluation methods, associated with each of the six dimensions, and synthesise guidelines for XAI evaluation, respectively, in the next six sections. 

\section{Fidelity to the Model}\label{sec:fidelity-slr}
We identify six criteria in the literature that are related to model fidelity. That is, these properties are used to examine how well an explanation captures the dynamics of the underlying predictive model. 

\begin{figure}[h!]
    \centering
    \includegraphics[scale=0.45]{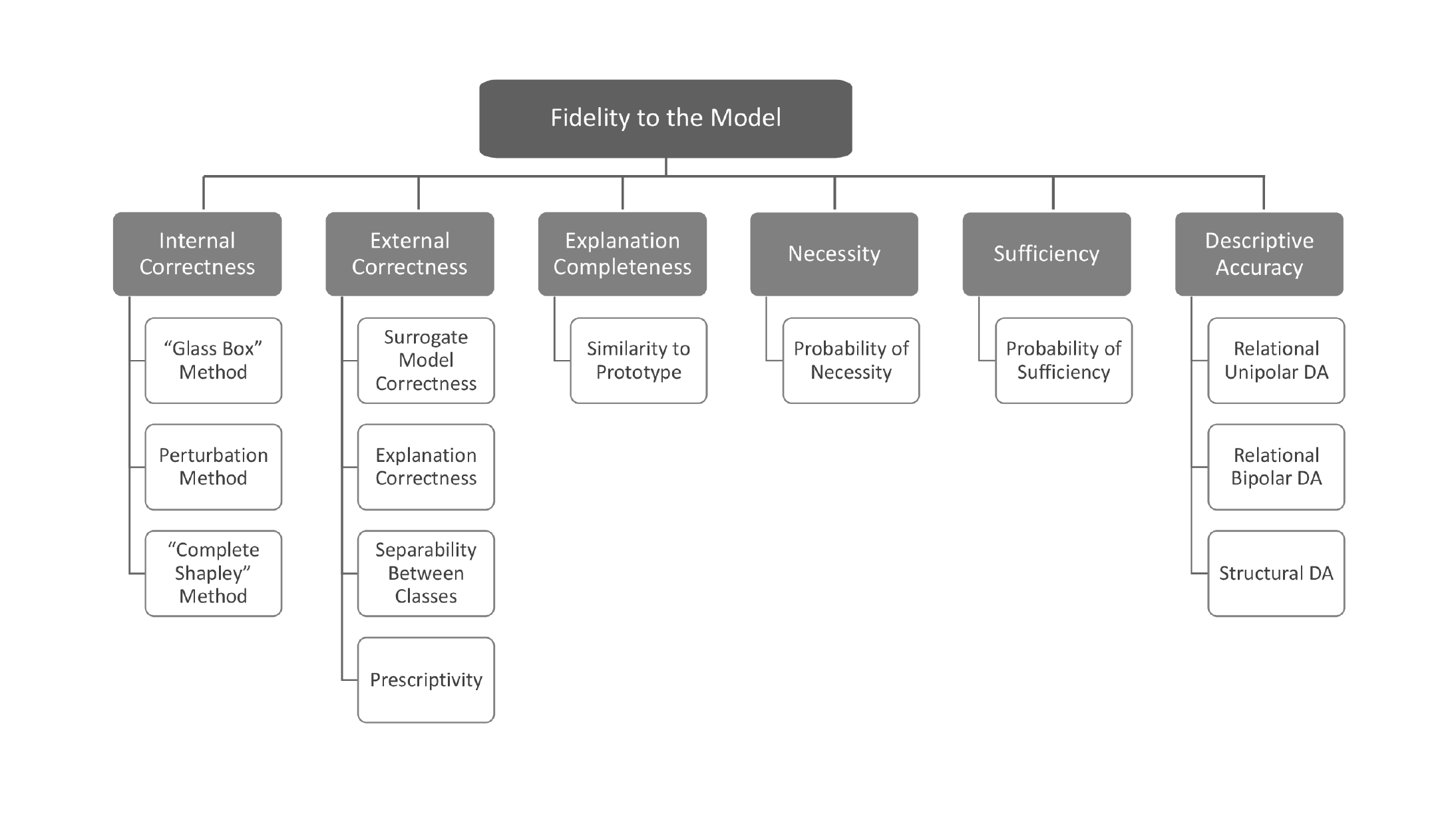}
    \vspace*{-3\baselineskip}
    \caption{The three levels of the taxonomy elaborating the explanation quality dimension of \emph{fidelity to the model}, the associated evaluation criteria, and the evaluation methods that may be used to test these explanation criteria. We identify at least one evaluation method for each criterion.}
    \label{fig:model-fid-taxonomy}
\end{figure}
\subsection{Internal Correctness of Explanations}\label{sec:internal-corr}
\vspace{.5\baselineskip}


\noindent 
\textbf{Evaluation Criterion.} 
Messalas, Kanellopoulos and Makris~\cite{Messalas2019} define two forms of fidelity, \textbf{internal fidelity} and \textbf{external fidelity}. We adopt their definitions and classify two forms of explanation correctness, the first of which is the internal correctness of explanations. This is the explanation's truthfulness with respect to the features that are used by the model, i.e., \emph{does the explanation correctly indicate feature importance to the model?} In the literature, this form of fidelity is also referred to as \textbf{fidelity}~\cite{Dai2022,Delaunay2022,Doumard2023,Ferrettini2021,Gaudel2022,Guidotti2021,Guidotti2022,Ozyegen2021,Zhou2019}, \textbf{correctness}~\cite{Cugny2022,Veerappa2022}, \textbf{soundness}~\cite{Maltbie2021}, \textbf{internal fidelity}~\cite{Velmurugan2021a}, \textbf{faithfulness}~\cite{Naretto2022}, \textbf{monotonicity}~\cite{Naretto2022,Stevens2022}, \textbf{truthfulness}~\cite{Mollas2022a}, \textbf{unipolar descriptive accuracy}~\cite{Albini2022} and \textbf{bipolar descriptive accuracy}~\cite{Albini2022}. Given the relatively diverse set of names specified in the literature, we propose \textbf{``internal correctness''} as the name for this criterion, for the reasons below:
\begin{enumerate}
    \item It reflects the internal fidelity name~\cite{Messalas2019,Velmurugan2021a}.
    \item It describes the expectation for the explanation, i.e. the explanation should be correct.
    \item It does not use terms that are used in the literature for more than one concept. For example, fidelity and faithfulness have often been used interchangeably to describe different forms of explanation completeness and explanation correctness.
\end{enumerate}

\noindent 
\textbf{Evaluation Methods.} Three methods are applied in the literature to measure the internal correctness of explanations:
\begin{itemize}
    \item \textit{``Glass box'' Method}: The ``glass box'' method compares explanations against the workings of a fully transparent and inherently interpretable, i.e. ``glass box'', predictive model~\cite{Albini2022,Dai2022,Gaudel2022,Guidotti2021,Guidotti2022,Jia2019,Stevens2022,Zhou2019}. This type of model offers an unambiguous truth of the model's decision-making. 
    \item \textit{Perturbation Method}: Given a data point and an explanation for that data point, this method perturbs the features in the data point and measures the impact of perturbation on the model prediction -- for an explanation to be correct, the features it deems important should produce a larger change than features it deems unimportant~\cite{Albini2022,Cugny2022,Dai2022,Maltbie2021,Mollas2022a,Ozyegen2021,Velmurugan2021a,Bobek2021a,Lundberg2020,Naretto2022,Ozyegen2021,Schlegel2021,Veerappa2022,Zhang2021}. 
    \item \textit{``Complete Shapley'' method}: This method measures well an explanation matches the explanation produced a different post hoc XAI technique, which is deemed ``definitive''~\cite{Doumard2023,Ferrettini2021,Letzgus2022,Duell2022,Loecher2022}. This method is typically applied when evaluating XAI techniques that are derived from another, for example, comparing optimisations of Shapley Value calculation against the original, full Shapley Value calculation~\cite{Doumard2023,Ferrettini2021,Letzgus2022}, which gives this evaluation method its name. 
\end{itemize}

\begin{table}[t!!!]
\caption{The settings under which it is appropriate (\ok) or not (\notok) to evaluate \textbf{Internal Correctness}, and any conditions associated with evaluation (\conditional).}
\label{tab:int-corr-guidelines}
\begin{tabular}{|p{0.13\linewidth}|p{0.15\linewidth}p{0.02\linewidth}p{0.21\linewidth}p{0.35\linewidth}|}
\hline
\multicolumn{1}{|l}{\textbf{Criteria}} & \multicolumn{1}{l}{\textbf{Category}} & \multicolumn{2}{c}{\textbf{Setting}} & \textbf{Notes} \\ \hline
\multirow{7}{\linewidth}{Internal Correctness} & \multirow{3}{\linewidth}{Evaluation Context} & \ok & General Purpose & \multirow{3}{\linewidth}{Evaluation methods may be limited under some settings for practicality.} \\ 
 &  & \conditional & Application-Based &  \\ 
 &  &  & & \\\cline{2-5} 
 & Data & \ok & All Forms of Tabular Data & Most appropriate evaluation method may depend on the data. \\ \cline{2-5} 
 & Models & \conditional & All Categories of Models & Of the three evaluation methods, the glass box evaluation method can only be applied if the model is fully transparent \\ \cline{2-5} 
 & \multirow{4}{\linewidth}{XAI techniques} & \ok & Feature Attribution & \multirow{4}{\linewidth}{Requires that the XAI technique indicate specific features as relevant to the prediction} \\ 
 &  & \notok & Feature Profiling &  \\ 
 &  & \conditional & Feature Sets &  \\ 
 &  & \ok & Rule-Based &  \\ \hline
\end{tabular}
\end{table}

\noindent 
\textbf{Evaluation Guidelines.} {We analyse the relevant context of existing evaluation methods (referred to as evaluation context), data, (predictive) model and XAI techniques, and then derive evaluation guidelines for assessing the internal correctness of explanations. These guidelines are summarised in Table~\ref{tab:int-corr-guidelines}.} 

\textit{Regarding Evaluation Context.} In the literature, this criterion is mostly evaluated in general-purpose settings. Two works conducting application-based evaluation use the perturbation approach~\cite{Ozyegen2021,Velmurugan2021a}. 
This lack of application-based evaluation highlights the limitations of all three evaluation methods:
\begin{itemize}
    \item ``Glass box'' evaluation: 
    This method of evaluation is not practical in real-world settings, as a glass box model does require post hoc explainability. 
    
    \item Perturbation-based evaluation: While this method is the most practical in real-world or application-based settings, surveyed literature suggests that the scale and type of perturbation affect evaluation results~\cite{Maltbie2021,Veerappa2022}. This may call into question the validity of the result, and this evaluation method requires further investigation.
    
    \item ``Complete Shapley'' evaluation: This method compares post hoc explanations against an explanation computed through an assumed ``definitive'' XAI technique. 
    This method is impractical for application-based evaluation as 1) if we consider a post hoc technique to be definitive, we do not require others; and 2) these ``definitive'' methods often require extensive computing resources or time. 
\end{itemize}

\textit{Regarding Data.} There are some differences with respect to the evaluation method applied for different forms of data. 
The complete Shapley method is not applied to multivariate time series data, likely because of the computational cost of explaining a high-dimensional dataset. The glass box method is applied when using univariate data~\cite{Gaudel2022}. Given that the perturbation method generally relies on perturbing individual features, it is unclear how this method may be applied for univariate data. 

\textit{Regarding Model.} There is some limitation on evaluation methods that may be applied to predictive models. When applying the glass box method, the model must be inherently interpretable. 
Models that the glass box method has been applied to include probabilistic~\cite{Albini2022}, linear-based~\cite{Dai2022,Gaudel2022,Guidotti2021,Guidotti2022}, rule-based~\cite{Guidotti2021,Guidotti2022} and tree-based~\cite{Jia2019,Stevens2022} -- the restriction is not related to a particular category of model but whether it is transparent. 

\textit{Regarding XAI Techniques.} There is some restriction on the XAI techniques to which this criterion is applicable. All surveyed works that evaluated internal correctness assess feature attribution or rule-based explanations. As internal correctness investigates whether a prediction is correctly attributed to a specific feature, or features, it cannot be evaluated for explanations that do not highlight specific features, such as profile-based explanations. 

{\textit{Other Considerations.} Works in the literature generally do not distinguish between XAI techniques' aims when evaluating fidelity. For example, while LIME indicates the importance of features locally to the model, SHAP values indicate a feature's contribution to the prediction, i.e. how much the feature shifts the value of the prediction from the average. However, existing evaluation methods generally seem to assess explanations in comparison to feature importance to the model. This is a research gap that future works must address.}

\subsection{External Correctness of Explanations}\label{sec:ext-corr-def}
\vspace{.5\baselineskip}

\noindent
\textbf{Evaluation Criterion.} Messalas, Kanellopoulos and Makris~\cite{Messalas2019} propose the criterion \textbf{external fidelity}, which we refer to in this work as external correctness. External correctness requires that the explanation can mimic the decision boundary of the black box model, i.e. that the model and the XAI technique produce the same predictions for the same data points.
Notably, the component that produces the prediction can vary across XAI techniques. 

In literature, this form of evaluation has many names. It is most commonly referred to as \textbf{fidelity}~\cite{Amparore2021,Balagopalan2022,Gatta2021,Gaudel2022,Guidotti2019a,Guidotti2022,Hatwell2020,Li2019,Maaroof2022,Naretto2022,Ranjbar2022,Shankaranarayana2019,Bjoerklund2022,Buonocore2022,Molnar2023,Moradi2021,Parimbelli2023}, \textbf{external fidelity}~\cite{Velmurugan2021a}, \textbf{exactness}~\cite{Zhou2019}, \textbf{adherence}~\cite{Delaunay2022}, \textbf{local concordance}~\cite{Amparore2021},  \textbf{prescriptivity}~\cite{Amparore2021}, \textbf{purity}~\cite{Gatta2021}, \textbf{precision}~\cite{Guidotti2022,Hatwell2020,Mollas2022,Mylonas2023}, \textbf{exclusive coverage}~\cite{Hatwell2020} and \textbf{hit}~\cite{Guidotti2019a,Maaroof2022}. Given the relatively diverse set of names proposed for this criterion in literature, we propose \textbf{``external correctness''} as the name for this criterion as it:
\begin{enumerate}
    \item reflects the name of external fidelity~\cite{Messalas2019};
    \item describes the expectation for the explanation, i.e. that it is correct; and
    \item does not use terms that may have been used in literature for more than one concept.
\end{enumerate}

\noindent
\textbf{Evaluation Methods.} This criterion may be measured in one of four ways:
\begin{itemize}
    \item \textit{Surrogate Model Correctness}: When testing a surrogate model, predictions are generated from both the original and surrogate models for a dataset (either the training set or a set of nearest neighbour points). Correctness is measured as the accuracy of the surrogate model with respect black box model's predictions~\cite{Balagopalan2022,Buonocore2022,Delaunay2022,Hatwell2020,Guidotti2022,Maaroof2022,Moradi2021,Naretto2022,Amparore2021,Bjoerklund2022,Chen2022,Li2019,Zhou2019,Gatta2021,Gaudel2022,Shankaranarayana2019,Afrabandpey2020}. 
    \item \textit{Explanation Correctness}: If an explanation indicates the most likely output value of the model, this likely output is treated as a prediction. For example, when evaluating PDP explanations, the value on the Y-axis of the PDP graph is used~\cite{Molnar2023}. When applied to decision rules, the result indicated by the rule is treated as a prediction, i.e. is the result true when the antecedent is true~\cite{Hatwell2020,Mollas2022,Mylonas2023,Gatta2021,Narodytska2019,Maaroof2022,Moradi2021,Guidotti2019a,Maaroof2022}. Explanation correctness is typically measured as the accuracy of the explanation with respect to the predictive model. 
    
    

    \item \textit{Separability Between Classes}: Some works measure external correctness as how easily data points of different classes can be distinguished using only explanations. These works generate a local explanation for all data points, fit (supervised or unsupervised) machine learning models to these explanations, then evaluate how well the model discriminates between data points of different classes~\cite{Gramegna2021,Arslan2022}. 

    \item \textit{Prescriptivity}: This method is proposed in the LEAF toolbox~\cite{Amparore2021}, and tests how similar the decision boundary of a surrogate model is to the decision boundary of the original model, by identifying the nearest counterfactual data point, i.e. the nearest data point for which the prediction ``flips''.
\end{itemize}

\noindent 
\textbf{Evaluation Guidelines.} {After analysing the literature, we find that there are few restrictions on evaluating the external correctness of explanations. We summarise these in Table~\ref{tab:ext-corr-guidelines}.} 

\textit{Regarding Evaluation Context.} In literature, external correctness is evaluated across both application-based and general-purpose contexts.

\textit{Regarding Data.} As external correctness evaluation methods require only the model prediction and the XAI technique, there is no limitation on data. 

\textit{Regarding Model.} Many types of models are used in external correctness evaluation in the literature. As the procedure for external correctness evaluation requires only a prediction from the model, there does not appear to be a limitation on the model.

\textit{Regarding XAI Techniques.} This criterion is only applicable to XAI techniques with certain characteristics. In the literature, all of the techniques evaluated produce explanations or explanation mechanisms that indicate the model output. External correctness evaluates how well an XAI technique approximates the model's decision boundary, and if the technique does not approximate this decision boundary, this criterion does not apply. 

\begin{table}[b!!!]
\caption{The settings under which it is appropriate (\ok) or not (\notok) to evaluate \textbf{External Correctness}, and any conditions associated with evaluation (\conditional).}
\label{tab:ext-corr-guidelines}
\begin{tabular}{|p{0.13\linewidth}|p{0.15\linewidth}p{0.02\linewidth}p{0.21\linewidth}p{0.35\linewidth}|}
\hline
\multicolumn{1}{|l}{\textbf{Criteria}} & \multicolumn{1}{l}{\textbf{Category}} & \multicolumn{2}{c}{\textbf{Setting}} & \textbf{Notes} \\ \hline
\multirow{4}{\linewidth}{External Correctness} & Evaluation Context & \ok & All Contexts &  \\ \cline{2-5} 
 & Data & \ok & All Forms of  Tabular Data &  \\ \cline{2-5} 
 & Models & \ok & All Categories of Models &  \\ \cline{2-5} 
 & XAI   techniques & \conditional & All XAI Techniques & 1) Explanation must indicate outcome or 2) XAI Technique must have a mechanism for prediction. \\ \hline
\end{tabular}
\end{table}

\subsection{Explanation Completeness}\label{sec:completeness}
\vspace{.5\baselineskip}

\noindent 
\textbf{Evaluation Criterion.} Explanation completeness is defined as how much of the predictive model behaviour is captured by the explanation~\cite{Cugny2022,Zhou2021}. This criterion is proposed and evaluated by the creators of the AutoXAI framework~\cite{Cugny2022}.
\vspace{.5\baselineskip}

\noindent
\textbf{Evaluation Methods.} Explanation completeness is measured as the average distance between local explanations for a dataset and explanations generated for a set of representative ``prototype'' data points. 
\vspace{.5\baselineskip}

\begin{table}[hb!]
\caption{The settings under which it is appropriate (\ok) or not (\notok) to evaluate \textbf{Explanation Completeness}, and any conditions associated with evaluation (\conditional).}
\label{tab:completeness-guidelines}
\begin{tabular}{|p{0.13\linewidth}|p{0.15\linewidth}p{0.02\linewidth}p{0.21\linewidth}p{0.35\linewidth}|}
\hline
\multicolumn{1}{|l}{\textbf{Criteria}} & \multicolumn{1}{l}{\textbf{Category}} & \multicolumn{2}{c}{\textbf{Setting}} & \textbf{Notes} \\ \hline
\multirow{4}{\linewidth}{Explanation Completeness} & Evaluation Context & \ok & All Contexts &  \\ \cline{2-5} 
 & Data & \conditional & All Forms of Tabular Data & Further investigation is required to determine suitability for tabular data \\ \cline{2-5} 
 & Models & \ok & All Categories of Models &  \\ \cline{2-5} 
 & XAI Techniques & \ok & All XAI Techniques &  \\ \hline
\end{tabular}
\end{table}

\noindent 
\textbf{Evaluation Guidelines.} Only one work in the surveyed literature~\cite{Cugny2022} evaluates explanation completeness. To extract the guidelines, we draw primarily from the details provided by Cugny et al~\cite{Cugny2022}. We summarise our guidelines in Table~\ref{tab:completeness-guidelines}. 

\textit{Regarding Evaluation Context.} Cugny et al~\cite{Cugny2022} provide an example use case, suggesting that explanation completeness can be evaluated in an application-based setting. Thus, there appear to be no restrictions on the context of the evaluation. 

\textit{Regarding Data.} It is unclear whether Cugny et al's~\cite{Cugny2022} evaluation method is appropriate when using tabular data. This evaluation method requires a set of ``representative'' data points. Cugny et al~\cite{Cugny2022} evaluate completeness in the natural language processing domain and use semantic similarity to identify representative data points. ``Representativeness'' is more ambiguous in tabular data, particularly if using all features. As a dataset's dimensionality increases, the neighbourhood of a data point also grows larger~\cite{Altman2018}. If only few features are used to choose representative data points, the choice of features may depend on the domain. For example, for a multivariate time series dataset, should representative data points be decided based on the time sequence, a single ``feature'' that changes over time, or a combination of the two? Further investigation is needed regarding this.

\textit{Regarding Model.} The model is not required for completeness evaluation, and does not restrict evaluation.

\textit{Regarding XAI Techniques.} There is no restriction associated with the XAI technique evaluated. All that is required is that two explanations produced by an XAI technique can be compared.
\subsection{Necessity}\label{sec:necessity}
\vspace{.5\baselineskip}

    
\noindent 
\textbf{Evaluation Criterion.} Watson et al~\cite{Watson2022} define necessity as a feature being necessary for the prediction outcome -- i.e. the feature is a necessary condition for the prediction if and only if the feature logically follows from the prediction. 
The authors note that this definition could also be applied to a group of features, rather than just one. 
For an explanation technique to meet necessity, it must be able to accurately capture features that are necessary for a prediction.
\vspace{.5\baselineskip}


\noindent
\textbf{Evaluation Methods.} Watson et al~\cite{Watson2022} calculate the probability of necessity given a single feature or set of features. That is, the probability of a feature or set of features being present, given a particular outcome. 
For feature attribution explanations, this is calculated using the top-K features in the explanation. 
\vspace{.5\baselineskip}

\noindent 
\textbf{Evaluation Guidelines.} Necessity is only measured in one work~\cite{Watson2022}. We rely on the details of use presented by Watson et al~\cite{Watson2022} to determine the appropriate evaluation of this criterion. We summarise the guidelines we have derived in Table~\ref{tab:necessity-guidelines}.

\textit{Regarding Evaluation Context.} Limitations posed by the context are hard to determine using only one study. Further investigation of the practical aspects of evaluation, e.g. time and computing resources needed, is necessary.

\textit{Regarding Data.} Only tabular datasets are used for evaluation by Watson et al~\cite{Watson2022}. As the calculation of the associated metric relies on individual features, it is unclear how this metric could be applied to time series data. 

\textit{Regarding Model.} The calculation of necessity requires only the model prediction, and the internal aspects of the model do not restrict it.

\textit{Regarding XAI Techniques.} This criterion is only relevant when evaluating explanations that indicate individual features as being relevant to the prediction. As such, it would not be appropriate when evaluating, for example, feature profiling explanations. 

\begin{table}[!htb]
\caption{The settings under which it is appropriate (\ok) or not (\notok) to evaluate \textbf{Necessity}, and any conditions associated with evaluation (\conditional).}
\label{tab:necessity-guidelines}
\begin{tabular}{|p{0.13\linewidth}|p{0.15\linewidth}p{0.02\linewidth}p{0.21\linewidth}p{0.35\linewidth}|}
\hline
\multicolumn{1}{|l}{\textbf{Criteria}} & \multicolumn{1}{l}{\textbf{Category}} & \multicolumn{2}{c}{\textbf{Setting}} & \textbf{Notes} \\ \hline
\multirow{11}{\linewidth}{Necessity} & \multirow{2}{\linewidth}{Evaluation Context} & \ok & General Purpose & \multirow{2}{\linewidth}{Needs investigation to determine practicality in real-world settings.} \\ 
 &  & \conditional & Application-Based &  \\ \cline{2-4} 
 & \multirow{4}{\linewidth}{Data} & \ok & Tabular & \multirow{4}{\linewidth}{Use in time series data requires further investigation} \\ 
 &  & \conditional & Multivariate Time Series &  \\ 
 &  & \conditional & Univariate Time Series &  \\ 
 &  & \ok & Synthetic &  \\ \cline{2-5} 
 & Models & \ok & All Categories of Models &  \\ \cline{2-5} 
 & \multirow{2}{\linewidth}{XAI techniques} & \ok & Feature Attribution &  \\ 
 &  & \notok & Feature Profiling &  \\ 
 &  & \ok & Feature Sets &  \\ 
 &  & \ok & Rule-Based &  \\ \hline
\end{tabular}
\end{table}
\subsection{Sufficiency}\label{sec:sufficiency}
\vspace{.5\baselineskip}


\noindent 
\textbf{Evaluation Criterion.} Watson et al~\cite{Watson2022} define sufficiency as a feature being sufficient for the prediction outcome with no other features influencing the prediction -- i.e. the feature is sufficient for the predicted outcome if and only if the outcome logically follows from the feature. 
This work notes that this definition could also be applied to a group of features, rather than just one. The authors specify that explanations should be able to capture features sufficient for the prediction. 
\vspace{.5\baselineskip}



\noindent
\textbf{Evaluation Methods.} Watson et al~\cite{Watson2022} calculate the probability of sufficiency given a single feature or set of features. That is, the probability of a model producing a particular prediction given feature or set of features.
For feature attribution explanations, this is calculated using the top-K features in the explanation. 
\vspace{.5\baselineskip}

\noindent 
\textbf{Evaluation Guidelines.}
Sufficiency is only measured in one work~\cite{Watson2022}. We rely on the details of use presented by Watson et al~\cite{Watson2022} to determine the appropriate evaluation of this criterion.  We summarise the guidelines we have derived in Table~\ref{tab:sufficiency-guidelines}.

\textit{Regarding Evaluation Context.} Limitations posed by the context are hard to determine using only one study. Further investigation of the practical aspects of evaluation, e.g. time and computing resources needed, is necessary.

\textit{Regarding Data.} Only tabular datasets are used for evaluation by Watson et al~\cite{Watson2022}. As the calculation of the associated metric relies on individual features, it is unclear how this metric could be applied to time series data. 

\textit{Regarding Model.} The calculation of necessity requires only the model prediction, and the internal aspects of the model do not restrict it.

\textit{Regarding XAI Techniques.} As with necessity, this criterion is only relevant when evaluating explanations that directly indicate individual features as being relevant to the prediction. 

\begin{table}[!htb]
\caption{The settings under which it is appropriate (\ok) or not (\notok) to evaluate \textbf{Sufficiency}, and any conditions associated with evaluation (\conditional).}
\label{tab:sufficiency-guidelines}
\begin{tabular}{|p{0.13\linewidth}|p{0.15\linewidth}p{0.02\linewidth}p{0.21\linewidth}p{0.35\linewidth}|}
\hline
\multicolumn{1}{|l}{\textbf{Criteria}} & \multicolumn{1}{l}{\textbf{Category}} & \multicolumn{2}{c}{\textbf{Setting}} & \textbf{Notes} \\ \hline
\multirow{12}{\linewidth}{Sufficiency} & \multirow{2}{\linewidth}{Evaluation Context} & \ok & General Purpose & \multirow{2}{\linewidth}{Needs investigation to determine
practicality in real-world settings.} \\ 
 &  & \conditional & Application-Based &  \\\cline{2-4} 
 & \multirow{4}{\linewidth}{Data} & \ok & Tabular & \multirow{4}{\linewidth}{Use in time series data requires further investigation} \\ 
 &  & \conditional & Multivariate Time Series &  \\ 
 &  & \conditional & Univariate Time Series &  \\ 
 &  & \ok & Synthetic &  \\ \cline{2-5} 
 & Models & \ok & All Categories of Models &  \\ \cline{2-5} 
 & \multirow{2}{\linewidth}{XAI techniques} & \ok & Feature Attribution &  \\ 
 &  & \notok & Feature Profiling &  \\ 
 &  & \ok & Feature Sets &  \\ 
 &  & \ok & Rule-Based &  \\ \hline
\end{tabular}
\end{table}
\subsection{Descriptive Accuracy}\label{sec:descriptive-accuracy}
\vspace{.5\baselineskip}

\noindent 
\textbf{Evaluation Criterion.} 
Albini et al~\cite{Albini2022} present a new criterion for evaluating model fidelity. This criterion is referred to as ``descriptive accuracy'' (DA) and 
assesses whether the explanation correctly identifies relationships between features, as captured by the model.
It is important to note that this form of evaluation assumes that all features in the dataset, as well as the target variable are discrete. 
\vspace{.5\baselineskip}


\noindent
\textbf{Evaluation Methods.} Three forms of evaluating DA are defined by Albini et al~\cite{Albini2022}:
\begin{itemize}
    \item \textit{Relational Unipolar DA.} If an explanation includes an unsigned, unidirectional relationship between two features (i.e. a relational unipolar explanation), this method tests whether change to the determining feature changes the value of the dependent feature~\cite{Albini2022}. 
    \item \textit{Relational Bipolar DA.} When a unidirectional, signed relationship between features is included in the explanation (i.e., a relational bipolar explanation), this method tests whether change to the determining variable changes the dependent variable in a manner consistent with the sign~\cite{Albini2022}. 
    \item \textit{Structural DA.} If a transparent predictive model includes a structural description of relationships between features in a dataset and the target variable, this method tests whether an explanation also contains those relationships~\cite{Albini2022}.  
\end{itemize}

\noindent 
\textbf{Evaluation Guidelines.}
Descriptive Accuracy is evaluated only in one work~\cite{Albini2022}. We rely on the details of use presented by Albini et al~\cite{Albini2022} to determine the appropriate evaluation of this criterion. We summarise the guidelines we have derived in Table~\ref{tab:da-guidelines}.

\textit{Regarding Evaluation Context.} 
The structural DA method is a form of ``glass box'' evaluation.
Therefore, as with the ``glass box'' method, structural DA is impractical to evaluate in application-based settings. 

\textit{Regarding Data.} Descriptions of the methods and metrics used by Albini et al~\cite{Albini2022} indicate that it should be applied to discrete variables. It is currently unclear how, or even whether, the metrics proposed by Albini et al~\cite{Albini2022} should be used with  datasets that have continuous variables. Additionally, given the emphasis of the evaluation method on individual features, it is unclear how applicable DA is to time series data. 

\textit{Regarding Model.} Only fully transparent predictive models with a structural description can be used when applying the structural DA method. 
When using either of the relational DA methods, models may be opaque, but must capture relationships between features, not just between features and the target variable.

\textit{Regarding XAI Techniques.} When using either of the relational DA methods of evaluation, the explanation must capture relationships between features, not just between features and the target variable. 

\begin{table}[!htb]
\caption{The settings under which it is appropriate (\ok) or not (\notok) to evaluate \textbf{Descriptive Accuracy}, and any conditions associated with evaluation (\conditional).}
\label{tab:da-guidelines}
\begin{tabular}{|p{0.13\linewidth}|p{0.15\linewidth}p{0.02\linewidth}p{0.21\linewidth}p{0.35\linewidth}|}
\hline
\multicolumn{1}{|l}{\textbf{Criteria}} & \multicolumn{1}{l}{\textbf{Category}} & \multicolumn{2}{c}{\textbf{Setting}} & \textbf{Notes} \\ \hline
\multirow{8}{\linewidth}{Descriptive Accuracy} & \multirow{2}{\linewidth}{Evaluation Context} & \ok & General Purpose & \multirow{2}{\linewidth}{Structural DA impractical in real-world settings.} \\ 
 &  & \conditional & Application-Based &  \\ \cline{2-5} 
 & \multirow{4}{\linewidth}{Data} & \ok & Tabular & \multirow{4}{\linewidth}{Use with continous variables and with time series data requires further investigation.} \\ 
 &  & \conditional & Multivariate Time Series &  \\
 &  & \conditional & Univariate Time Series &  \\ 
 &  & \ok & Synthetic &  \\ \cline{2-5} 
 & Models & \conditional & All Categories of Models & Structural DA: Fully transparent model only. Relational DA: Model should capture relationships between features. \\ \cline{2-5} 
 & XAI   techniques & \conditional & All XAI Techniques & Relational DA: Explanation should include relationships between features. \\ \hline
\end{tabular}
\end{table}

\section{Explanation Invariance}\label{sec:invariance-slr}
We identified four criteria related to the variance and invariance of explanations in the literature. Three of these are related to how explanations change across a dataset. The final criterion is related to how the characteristics of the underlying model affect the content of the explanation.

\begin{figure}[!htb]
    \centering
    \includegraphics[scale=0.5]{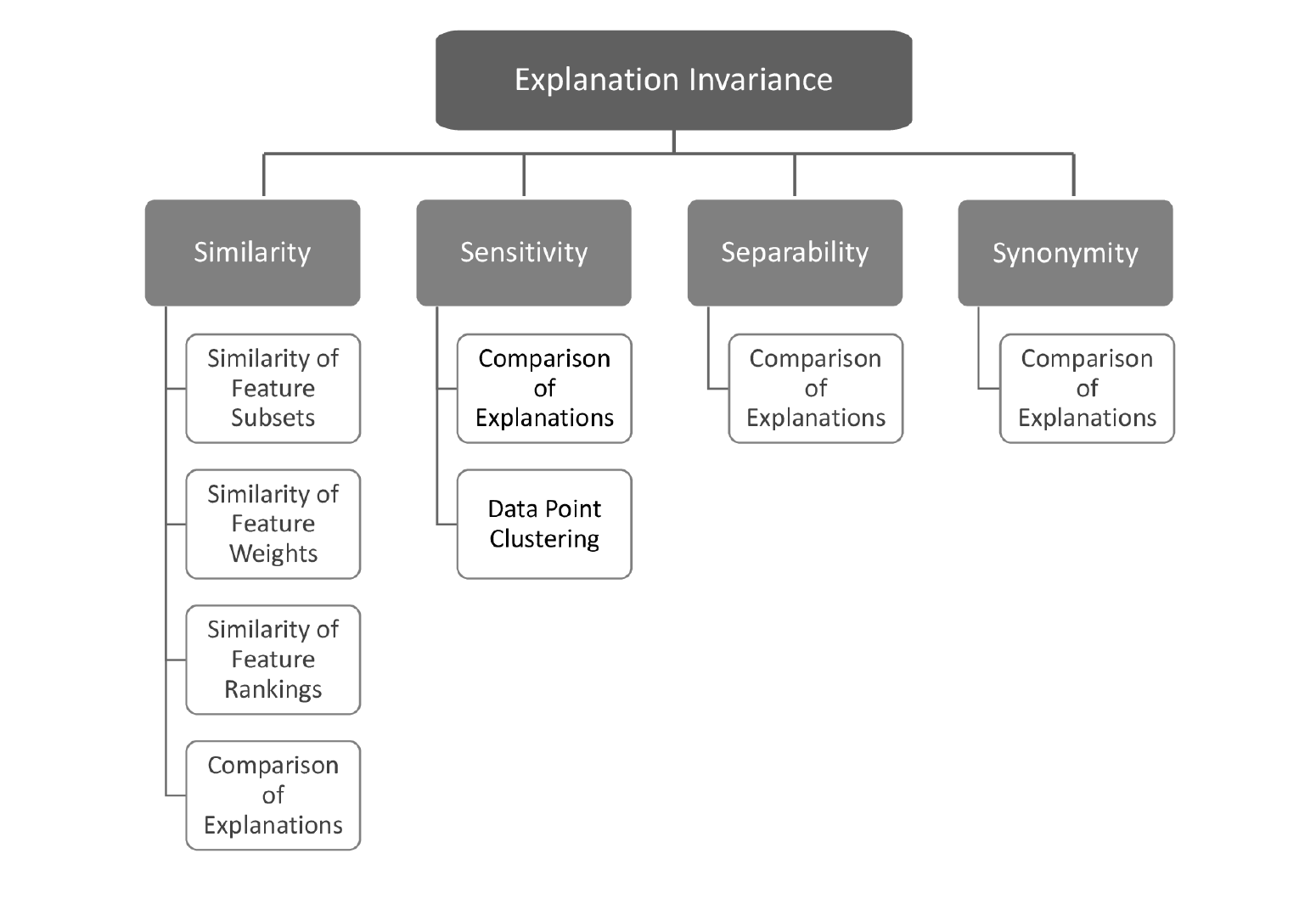}
    \caption{The three levels of the taxonomy elaborating the explanation quality dimension of \emph{explanation invariance}, the associated evaluation criteria, and the evaluation methods that may be used to test these explanation criteria. We identify at least one evaluation method for each criterion.}
    \label{fig:invariance-taxonomy}
\end{figure}

\subsection{Similarity}\label{sec:similarity}
\vspace{.5\baselineskip}

\noindent
\textbf{Evaluation Criterion.} Similarity measures the consistency of explanations generated for the same data point and predictive model by a single XAI technique, i.e. whether the XAI technique always produces the same explanations for a given data point.
Given that randomness is an inherent part of the explanation generation process in some techniques, e.g. LIME, it is necessary to understand to what degree this randomness affects explanations. 
This criterion is known by multiple names in literature. It is referred to as \textbf{stability}~\cite{Mueller2022,Ranjbar2022,Saini2022, Shankaranarayana2019,Velmurugan2021b}, \textbf{identity}~\cite{Buonocore2022,Ferraro2022,Hailemariam2020,Parimbelli2023,Guidotti2019a,Guidotti2022}, \textbf{variance}~\cite{Mollas2022} and \textbf{reiteration similarity}~\cite{Amparore2021}. Although identity is the most common name applied to this criterion, it is not fully descriptive of this criterion. Stability, another common name in the literature, is also often used to refer to the criterion of sensitivity (see Section~\ref{sec:sensitivity-def}). As such, we choose the name \textbf{similarity}, as it clearly describes this criterion, i.e. how similar are explanations for the same data point.
\vspace{.5\baselineskip}

\noindent
\textbf{Evaluation Methods.} There are four methods of evaluating similarity. All methods produce multiple explanations for a single data point, and evaluate similarity as the consistency of explanations across this set of explanations: 
\begin{itemize}
    \item \textit{Similarity of Feature Subsets}: Most works evaluate the similarity of the subset of features deemed most important by the explanation, i.e. how much does this subset change across explanations~\cite{Amparore2021,Saini2022,Velmurugan2021b,Chen2022,Ranjbar2022,Mollas2022,Guidotti2019a}. 
    \item \textit{Similarity of Feature Weights}: This method measures the variation in feature weights for feature attribution explanations~\cite{Shankaranarayana2019,Velmurugan2021b,Dai2022,Guidotti2022}.
    \item \textit{Similarity of Feature Rankings}: Müller et al~\cite{Mueller2022} evaluate the similarity of feature ranks in feature attribution, rather than weights.
    \item \textit{Comparison of Explanations}: Several works check the percentage of data points for which an XAI technique produces identical explanations for the same data point across repeated runs~\cite{Buonocore2022,Ferraro2022,Hailemariam2020}.
\end{itemize}

\noindent 
\textbf{Evaluation Guidelines.} {We analyse the relevant context of existing evaluation methods (referred to as evaluation context), data, (predictive) model and XAI techniques, and then derive evaluation guidelines for assessing explanation similarity. We summarise these guidelines in Table~\ref{tab:similarity-guidelines}.} 

\textit{Regarding Evaluation Context.} Similarity is evaluated across many contexts in literature. Context does not appear to limit evaluation. 

\textit{Regarding Data.} Evaluation methods examine only the explanation. There is no limitation associated with the data used to train the model.

\textit{Regarding Model.} Evaluation methods examine only the explanation. There is no limitation associated with the predictive model being explained.

\textit{Regarding XAI Techniques.} Only feature attribution and rule-based explanations are evaluated for similarity in the literature. However, as long as two explanations can definitively be said to be identical or not, similarity can be evaluated.

\textit{Other Considerations.} We suggest that similarity is the most fundamental explanation invariance criterion. Similarity requires that an XAI technique consistently produce the same explanation for a single data point. Other criteria typically extend from this requirement, e.g. that change in explanations across a neighbourhood are consistent. 
If an XAI technique cannot satisfy similarity, other forms of consistency are unlikely to be present.

\begin{table}[!htb]
\caption{The settings under which it is appropriate (\ok) or not (\notok) to evaluate \textbf{Explanation Similarity}, and any conditions associated with evaluation (\conditional).}
\label{tab:similarity-guidelines}
\begin{tabular}{|p{0.13\linewidth}|p{0.15\linewidth}p{0.02\linewidth}p{0.21\linewidth}p{0.35\linewidth}|}
\hline
\multicolumn{1}{|l}{\textbf{Criteria}} & \multicolumn{1}{l}{\textbf{Category}} & \multicolumn{2}{c}{\textbf{Setting}} & \textbf{Notes} \\ \hline
\multirow{4}{\linewidth}{Similarity} & Evaluation Context & \ok & All Contexts &  \\ \cline{2-5} 
 & Data & \ok & All Forms of Tabular Data &  \\ \cline{2-5} 
 & Models & \ok & All Categories of Models &  \\ \cline{2-5} 
 & XAI Techniques & \ok & All XAI Techniques &  \\ \hline 
\end{tabular}
\end{table}
\subsection{Sensitivity}\label{sec:sensitivity-def}
\vspace{0.5\baselineskip}

\noindent 
\textbf{Evaluation Criterion.} Sensitivity refers to the continuity of explanations across the different permutations of the data, such that small perturbations to the data do not cause large changes in explanation and vice versa. 
This criterion is also known as \textbf{robustness}~\cite{Botari2020,Doumard2023,Jakubowski2022}, \textbf{stability}~\cite{Bobek2021,Bobek2021a,Dai2022,Ferraro2022,Guidotti2022,Hailemariam2020,Naretto2022}, \textbf{similarity}~\cite{Parimbelli2023}, \textbf{coherence}~\cite{Guidotti2022}, \textbf{continuity}~\cite{Cugny2022} and \textbf{consistency}~\cite{Tritscher2023}. This criterion is most commonly referred to as ``robustness'' or ``stability'' in the literature, but both of these terms are also often applied to other forms of evaluation, i.e. robustness against adversarial attacks (Section~\ref{sec:adversarial-robustness-def}), and the consistency of explanations for a single data point (Section~\ref{sec:similarity}). Other names given to this criterion are also ambiguous or used to refer to other qualities of explanations. Thus, we choose the name \textbf{``sensitivity''}, as it:
\begin{enumerate}
    \item to the best of our knowledge, is not used to refer to other aspects of explainability;
    \item is descriptive of the criterion, aka how sensitive explanations are to changes in the data; and
    \item is synonymous with common names applied to this criterion, such as ``robustness'' and ``stability''.
\end{enumerate}

\noindent
\textbf{Evaluation Methods.} Two methods of explanation evaluation are typically applied to measure explanation sensitivity:
\begin{itemize}
    \item \textit{Comparison of Explanations}: This approach directly compares two explanations to assess whether the difference between explanations is proportional to the difference between the data points they explain~\cite{Bobek2021a,Bobek2021,Cugny2022,Doumard2023,Guidotti2022,Jakubowski2022,Naretto2022,Dai2022,Botari2020,Hailemariam2020,Tritscher2023,Dai2022,Botari2020}.
    \item \textit{Data Point Clustering}: This approach clusters all data points in the training set based on the feature weights provided by feature attribution explanations, and calculates the percentage of data points for which the expected and actual classes match~\cite{Ferraro2022}. This method is similar to the ``separability between classes'' method of external correctness evaluation (see Section~\ref{sec:ext-corr-def}).
\end{itemize}

\noindent 
\textbf{Evaluation Guidelines.} {After analysing the literature, we find few restrictions associated with the evaluation of explanation sensitivity. We summarise our guidelines in Table~\ref{tab:sensitivity-guidelines}.} 

\textit{Regarding Evaluation Context.} Explanation sensitivity is evaluated in literature across both contexts. However, the results are highly dependent on individual components of the prediction pipeline, and evaluation in benchmark settings may not produce useful results~\cite{Doumard2023}.

\textit{Regarding Data.} Evaluation requires only that two data points can be quantitatively compared. As such, there are no limitations on the data. 

\textit{Regarding Model.} The model is not directly used in evaluation, and so choice of model does not restrict sensitivity evaluation.

\textit{Regarding XAI Techniques.} In literature, sensitivity is evaluated only for feature attribution and rule-based explanations. However, there should be no limitation on sensitivity evaluation as long as two explanations can be quantitatively compared.

\textit{Other Considerations.} The data point clustering method of evaluation requires further investigation. This method is almost identical to the class separability method of external correctness evaluation, and it is unclear exactly how the results of the two criteria may be related.

\begin{table}[!htb]
\caption{The settings under which it is appropriate (\ok) or not (\notok) to evaluate \textbf{Explanation Sensitivity}, and any conditions associated with evaluation (\conditional).}
\label{tab:sensitivity-guidelines}
\begin{tabular}{|p{0.13\linewidth}|p{0.15\linewidth}p{0.02\linewidth}p{0.21\linewidth}p{0.35\linewidth}|}
\hline
\multicolumn{1}{|l}{\textbf{Criteria}} & \multicolumn{1}{l}{\textbf{Category}} & \multicolumn{2}{c}{\textbf{Setting}} & \textbf{Notes} \\ \hline
\multirow{6}{\linewidth}{Sensitivity} & \multirow{3}{\linewidth}{Evaluation Context} & \conditional & General Purpose & \multirow{6}{\linewidth}{Usefulness of results in benchmark settings may be variable. More investigation is also needed for ``Clustering'' method of evaluation to determine suitability.} \\
 &  & \ok & Application-Based &  \\ 
 &  &  & &  \\ \cline{2-4} 
 & Data & \ok & All Forms of Tabular Data &  \\ \cline{2-4} 
 & Models & \ok & All Categories of Models &  \\ \cline{2-4} 
 & XAI Techniques & \ok & All XAI Techniques &  \\ \hline
\end{tabular}
\end{table}

\subsection{Separability}
\vspace{.5\baselineskip}

\noindent 
\textbf{Evaluation Criterion.} Separability of explanations requires that different data points should have different explanations. This is typically evaluated for feature attribution explanations, i.e. where feature weights are assigned for each feature by the explanation. This form of explanation is expected to be highly local.
\vspace{.5\baselineskip}

\noindent
\textbf{Evaluation Methods.} Separability is evaluated by comparing explanations for two different data points, to determine whether the produced explanations are identical. Separability is typically measured as the percentage of data points for which the generated explanations are not unique~\cite{Buonocore2022,Ferraro2022,Hailemariam2020}. 
\vspace{.5\baselineskip}

\noindent 
\textbf{Evaluation Guidelines.} {We analyse the relevant context of existing evaluation methods (referred to as evaluation context), data, (predictive) model and XAI techniques, and then derive evaluation guidelines for assessing explanation separability. We summarise these in Table~\ref{tab:separability-guidelines}.} 

\textit{Regarding Evaluation Context.} Separability is evaluated in both contexts in literature. Context does not appear to limit evaluation.

\textit{Regarding Data.} Separability evaluation is based on explanations for two different data points. As long as data points from the same dataset can be definitively said to be identical or different, separability can be evaluated.

\textit{Regarding Model.} The model is not directly used in evaluation, and choice of predictive model does not limit evaluation.

\textit{Regarding XAI Techniques.} Separability is only applicable to feature attribution explanations. Feature attribution explanations are highly local, often explaining only a single data point.
However, for other forms of explanations, ``local'' typically refers to a data neighbourhood. For example, rule-based explanations indicate a region of data that produces a specific result. 
As such, explanations other than feature attribution are not expected to be unique for every data point.

\begin{table}[!htb]
\caption{The settings under which it is appropriate (\ok) or not (\notok) to evaluate \textbf{Explanation Separability}, and any conditions associated with evaluation (\conditional).}
\label{tab:separability-guidelines}
\begin{tabular}{|p{0.13\linewidth}|p{0.15\linewidth}p{0.02\linewidth}p{0.21\linewidth}p{0.35\linewidth}|}
\hline
\multicolumn{1}{|l}{\textbf{Criteria}} & \multicolumn{1}{l}{\textbf{Category}} & \multicolumn{2}{c}{\textbf{Setting}} & \textbf{Notes} \\ \hline
\multirow{7}{\linewidth}{Separability} & Evaluation Context & \ok & All Contexts &  \\ \cline{2-5} 
 & Data & \ok & All Forms of Tabular Data &  \\ \cline{2-5} 
 & Models & \ok & All Categories of Models &  \\ \cline{2-5} 
 & \multirow{4}{\linewidth}{XAI Techniques} & \ok & Feature Attribution & \multirow{4}{*}{} \\
 &  & \notok & Feature Profiling &  \\
 &  & \notok & Feature Sets &  \\
 &  & \notok & Rule-Based &  \\ \hline
\end{tabular}
\end{table}
\subsection{Synonymity}
\vspace{0.5\baselineskip}

\noindent
\textbf{Evaluation Criterion.}
Synonymity refers to the consistency of explanations for two machine learning models, which are trained on the same dataset, and which have similar decision boundaries. These synonymous models are expected to have similar or identical explanations. Adherence to synonymity indicates that the explanation is founded on the phenomenon being captured by the model, rather than some characteristic of the model.
This criterion is typically called \textbf{consistency}~\cite{Bobek2021,Bobek2021a,Jakubowski2022}. However, consistency may also refer to a property specific to SHAP~\cite{Lundberg2020,Lundberg2017}, and is often used as alternative names or descriptions for other explanation invariance criteria. As such, we instead name this criterion \textbf{``synonymity''} -- that is, when models are synonymous, explanations for those models must similarly be the same. 
\vspace{0.5\baselineskip}


\noindent
\textbf{Evaluation Methods.} Synonymity is measured using two different machine learning models which are trained on identical data and make similar or identical predictions. Synonymity evaluation calculates the similarity of explanations for the same data point for each model~\cite{Bobek2021a,Jakubowski2022}. 
\vspace{0.5\baselineskip}

\noindent 
\textbf{Evaluation Guidelines.} {We analyse the evaluation context, data, predictive model and XAI techniques, and then derive evaluation guidelines for assessing explanation synonymity. We summarise these guidelines in Table~\ref{tab:synonymity-guidelines}.} 

\textit{Regarding Evaluation Context.} In literature, synonymity evaluation is not limited by context. However, it requires two trained models for evaluation, which may be impractical in real-world settings.

\textit{Regarding Data.} Data is not directly used in the evaluation. Synonymity can be evaluated as long as two data points can be definitively said to be identical.

\textit{Regarding Model.} No predictive models are directly used in the evaluation of explanation synonymity. Evaluation of synonymity only requires two predictive models and which are trained on the same dataset and make the same predictions. 

\textit{Regarding XAI Techniques.} Evaluation of synonymity requires that two explanations produced by the same technique can be said to be identical or not. As such, there is no limitation on the type of explanation that may be assessed.

\begin{table}[!htb]
\caption{The settings under which it is appropriate (\ok) or not (\notok) to evaluate \textbf{Explanation Synonymity}, and any conditions associated with evaluation (\conditional).}
\label{tab:synonymity-guidelines}
\begin{tabular}{|p{0.13\linewidth}|p{0.15\linewidth}p{0.02\linewidth}p{0.21\linewidth}p{0.35\linewidth}|}
\hline
\multicolumn{1}{|l}{\textbf{Criteria}} & \multicolumn{1}{l}{\textbf{Category}} & \multicolumn{2}{c}{\textbf{Setting}} & \textbf{Notes} \\ \hline
\multirow{5}{\linewidth}{Synonymity} & \multirow{2}{\linewidth}{Evaluation Context} & \ok & General Purpose & \multirow{2}{\linewidth}{Further investigation needed regarding practicality} \\
 &  & \conditional & Application-Based & \\ \cline{2-5} 
 & Data & \ok & All Forms of Tabular Data &  \\ \cline{2-5} 
 & Models & \ok & All Categories of Models &  \\ \cline{2-5} 
 & XAI Techniques & \ok & All XAI Techniques &  \\ \hline 
\end{tabular}
\end{table}

\section{Fidelity to the Ground Truth}\label{sec:data-fid-slr}
Many works in literature evaluate the fidelity of explanations by checking its compliance to the training dataset or the process that produced this dataset, rather than the model. Thus, we distinguish these criteria from the evaluation of explanation fidelity to the model. 

\begin{figure}[!htb]
    \centering
    \includegraphics[scale=0.45]{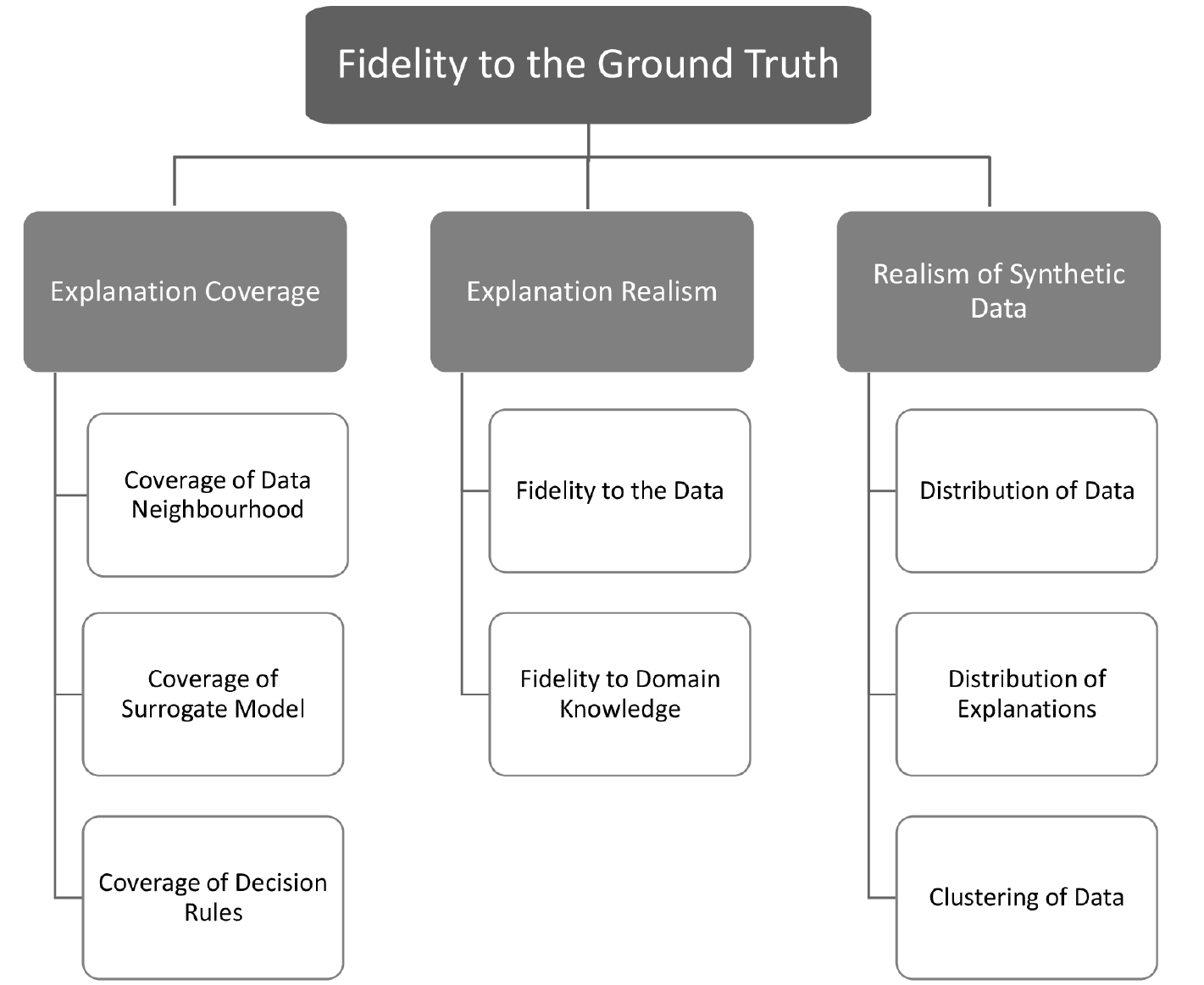}
    \caption{The three levels of the taxonomy elaborating the explanation quality dimension of \emph{fidelity to the ground truth}, the associated evaluation criteria, and the evaluation methods that may be used to test these explanation criteria. We identify at least one evaluation method for each criterion.}
    \label{fig:truth-fid-taxonomy}
\end{figure}

\subsection{Explanation Coverage}\label{sec:coverage-slr-def}
\vspace{0.5\baselineskip}

\noindent
\textbf{Evaluation Criterion.} Based on existing literature, we propose that coverage of explanations be defined as how much of the data space is covered by the explanation.
Coverage may be measured across the training set, or only across the neighbourhood of the data point being explained. 
\vspace{0.5\baselineskip}

\noindent
\textbf{Evaluation Methods.} We identified three different methods of evaluating coverage, each at different steps of explanation generation:
\begin{itemize}
    \item \textit{Coverage of Data Neighbourhood}: Coverage may be measured when clustering data points to identify the neighbourhood of the data point being explained. La Gatta et al~\cite{Gatta2021} measure the coverage of each cluster relative to the training set.

    \item \textit{Coverage of Surrogate Model}: Björklund, Mäkelä and Puolamäki~\cite{Bjoerklund2022} evaluate the coverage of local surrogate models across the original dataset by determining the proportion of data points in the training set for which the surrogate model meets a certain level of accuracy. 

    \item \textit{Coverage of Decision Rules}: Coverage of decision rules is typically measured as the fraction of some dataset, either a training set or a data neighbourhood, for which the rule applies, i.e. the fraction of the dataset that meets that antecedents of the rule~\cite{Delaunay2022,Hatwell2020,Gatta2021,Guidotti2019a,Guidotti2022,Mollas2022,Moradi2021,Mylonas2023}.
\end{itemize}

\noindent 
\textbf{Evaluation Guidelines.} {We analyse the literature and derive evaluation guidelines for assessing explanation coverage. This criterion has several restrictions associated with evaluation. We summarise these in Table~\ref{tab:coverage-guidelines}.} 

\textit{Regarding Evaluation Context.} In the literature, explanation coverage is only evaluated in general-purpose contexts. It's unclear from examination of the evaluation method why this is, and warrants further investigation.

\textit{Regarding Data.} In the literature, this criterion is only evaluated using standard tabular data. More investigation is needed regarding evaluation using time series data, particularly if coverage of surrogate models evaluated, which relies on selection of representative data points. As noted in Section~\ref{sec:completeness}, selection of representative data points may be ambiguous.

\textit{Regarding Model.} The model is not directly used in evaluation of explanation coverage, and does not limit evaluation.

\textit{Regarding XAI Techniques.} Explanation coverage is evaluated for both decision rules and feature attribution explanations in the literature. However, evaluation of feature attribution explanations requires more investigation, as feature attribution explanations are extremely local, and typically explain a single data point. It is unclear whether evaluating feature attribution explanation coverage is useful and relevant. Relevance of coverage to other types of explanations also requires investigation.

\begin{table}[!htb]
\caption{The settings under which it is appropriate (\ok) or not (\notok) to evaluate \textbf{Explanation Coverage}, and any conditions associated with evaluation (\conditional).}
\label{tab:coverage-guidelines}
\begin{tabular}{|p{0.13\linewidth}|p{0.15\linewidth}p{0.02\linewidth}p{0.21\linewidth}p{0.35\linewidth}|}
\hline
\multicolumn{1}{|l}{\textbf{Criteria}} & \multicolumn{1}{l}{\textbf{Category}} & \multicolumn{2}{c}{\textbf{Setting}} & \textbf{Notes} \\ \hline
\multirow{12}{\linewidth}{Explanation Coverage} & \multirow{2}{\linewidth}{Evaluation Context} & \ok & General Purpose & \multirow{2}{\linewidth}{Usefulness in application-based contexts requires investigation} \\
 &  & \conditional & Application-Based &  \\\cline{2-5} 
 & \multirow{4}{\linewidth}{Data} & \ok & Tabular & \multirow{4}{\linewidth}{Further investigation needed regarding evaluation on time series data} \\
 &  & \conditional & Multivariate Time Series &  \\
 &  & \conditional & Univariate Time Series &  \\
 &  & \ok & Synthetic &  \\ \cline{2-5} 
 & Models & \ok & All Categories of Models &  \\ \cline{2-5} 
 & \multirow{4}{\linewidth}{XAI techniques} & \conditional & Feature Attribution & \multirow{4}{\linewidth}{Further investigation needed regarding evaluation for XAI techniques that are not rule-based} \\
 &  & \conditional & Feature Profiling &  \\
 &  & \conditional & Feature Sets &  \\
 &  & \ok & Rule-Based &  \\ \hline
\end{tabular}
\end{table}
\subsection{Explanation Realism}
\vspace{0.5\baselineskip}

\noindent
\textbf{Evaluation Criterion.} We propose that explanation realism be defined as the degree to which an explanation reflects the true reality of a domain.
This criterion is often referred to as \textbf{fidelity}~\cite{Jia2019,Mueller2022}, \textbf{faithfulness}~\cite{Tritscher2020} and \textbf{precision}~\cite{Moscato2021}. When applied to anomaly detection tasks, this criterion may be referred to as \textbf{blame attribution error}~\cite{Sipple2022}. As the first three names may also refer to fidelity to the model, and the latter name is task-specific, we choose the name \textbf{explanation realism}. This name accurately describes what this criterion measures -- how correct explanations are with respect to real-life phenomena.
\vspace{0.5\baselineskip}

\noindent
\textbf{Evaluation Methods.} Methods evaluating explanation realism select a set of ``true features'' and a set of explanation features and measure explanation realism as the similarity between the two sets. The primary difference between evaluation methods is how the set of true features is identified:
\begin{itemize}
    \item \textit{Fidelity to the Data}: This method is sometimes treated as a form of ``glass box'' fidelity evaluation. Models are trained on extremely simple tabular datasets for which the ground truth is already known, and knowledge of the data to determine the true features~\cite{Jia2019,Moscato2021,Tritscher2020}.

    \item \textit{Fidelity to Domain Knowledge}: This method relies on expert knowledge to identify the set of true features, typically by using datasets annotated by domain experts~\cite{Mueller2022,Sipple2022,Tritscher2023}. 
\end{itemize}

\noindent 
\textbf{Evaluation Guidelines.} {We analyse the literature and derive evaluation guidelines for assessing explanation realism. These are summarised in Table~\ref{tab:explanation-realism-guidelines}.} 

\textit{Regarding Evaluation Context.} In the surveyed literature, this criterion is evaluated in both general-purpose and application-based settings. However, evaluation is impractical in a real-world setting. The ``fidelity to the data'' method has the same weaknesses as the ``glass box'' method of internal correctness evaluation (see Section~\ref{sec:internal-corr}). The second evaluation method uses domain knowledge to test an explanation's validity. If the explanation does not match the domain knowledge, there are two possibilities: the explanation is faithful to the model, and the model does not reflect reality; or the explanation is not faithful to the model, which does reflect reality. If we know, or assume, that the explanation is faithful to the model, then this evaluates the model not the explanation. If explanation fidelity is unclear, then comparing the explanation to domain knowledge is not informative.

\textit{Regarding Data.} Evaluation of this criterion relies on knowledge of ``importance'' from the data or the domain. As such, the data does not limit the evaluation of explanation realism.

\textit{Regarding Model.} The model is not directly required in the evaluation of explanation realism. Choice of model does not limit evaluation.

\textit{Regarding XAI Techniques.} Both feature attribution and rule-based techniques are evaluated in the literature. This criterion is likely applicable to any technique that indicates specific features as important to the prediction. However, further investigation is needed in this regard.

\begin{table}[!htb]
\caption{The settings under which it is appropriate (\ok) or not (\notok) to evaluate \textbf{Explanation Realism}, and any conditions associated with evaluation (\conditional).}
\label{tab:explanation-realism-guidelines}
\begin{tabular}{|p{0.13\linewidth}|p{0.15\linewidth}p{0.02\linewidth}p{0.21\linewidth}p{0.35\linewidth}|}
\hline
\multicolumn{1}{|l}{\textbf{Criteria}} & \multicolumn{1}{l}{\textbf{Category}} & \multicolumn{2}{c}{\textbf{Setting}} & \textbf{Notes} \\ \hline
\multirow{8}{\linewidth}{Explanation Realism} & \multirow{2}{\linewidth}{Evaluation Context} & \ok & General Purpose &  \\
 &  & \notok & Application-Based &  \\ \cline{2-5} 
 & Data & \ok & All Forms of Tabular Data &  \\ \cline{2-5} 
 & Models & \ok & All Categories of Models &  \\ \cline{2-5} 
 & \multirow{4}{\linewidth}{XAI techniques} & \ok & Feature Attribution & \multirow{4}{\linewidth}{Further investigation needed to determine applicability for feature profiling and feature set explanations} \\
 &  & \conditional & Feature Profiling &  \\
 &  & \conditional & Feature Sets &  \\
 &  & \ok & Rule-Based &  \\ \hline
\end{tabular}
\end{table}
\subsection{Realism of Synthetic Data}\label{sec:data-realism}
\vspace{0.5\baselineskip}

\noindent
\textbf{Evaluation Criterion.} To calculate a local explanation, many XAI techniques generate a synthetic dataset to represent the local neighbourhood of the data point being explained. We propose that realism of synthetic data be defined as the degree to which this synthetic dataset reflects the distributions in the training set.
In the literature, this criterion is referred to as \textbf{data fidelity}~\cite{Molnar2023}, or \textbf{silhouette}~\cite{Guidotti2022} and \textbf{cluster purity}~\cite{Bjoerklund2022}. Given that ``fidelity'' is an ambiguous term in XAI evaluation, and that the other two names are associated with specific metrics, we choose a new name for this criterion. We choose \textbf{``realism of synthetic data''}, as it unambiguously specifies what is being evaluated.
\vspace{0.5\baselineskip}

\noindent
\textbf{Evaluation Methods.} There are three evaluations methods that may be applied when assessing realism of synthetic data:
\begin{itemize}
    \item \textit{Distribution of Data}: This method directly compares the synthetic dataset to the training dataset, using metrics such as the silhouette score~\cite{Guidotti2022} and the mean maximum discrepancy measure~\cite{Molnar2023}. 

    \item \textit{Distribution of Explanations}: This method determines compares the distribution of feature weights in feature attribution explanations for the original and synthetic datasets, to see how well they match~\cite{Saini2022}.

    \item \textit{Clustering of Data}: This method is applied to XAI techniques that use dimension reduction, rather than generating a synthetic neighbourhood, and aims to assess whether the original neighbourhoods in the data are maintained after dimension reduction. The original and reduced data are clustered, and the similarity between clusters is evaluated~\cite{Bjoerklund2022}.
\end{itemize}

\noindent 
\textbf{Evaluation Guidelines.} {We analyse the literature and derive evaluation guidelines for assessing the realism of synthetic data. These are summarised in Table~\ref{tab:data-realism-guidelines}.} 

\textit{Regarding Evaluation Context.} This criterion assesses a single component of the explanation generation process, and is only evaluated in the surveyed literature when testing new XAI techniques or extensions to XAI techniques. 

\textit{Regarding Data.} It is unclear exactly how the three evaluation methods are applicable to time series data. Further investigation is needed.

\textit{Regarding Model.} As with other criteria in this dimension, the model is not directly used in evaluation. The choice of predictive model does not limit evaluation of this criterion.

\textit{Regarding XAI Techniques.} This criterion only applies to XAI techniques that produce a synthetic neighbourhood as part of the explanation generation, such as LIME. Model-specific techniques, for example, typically do not include this step, and so this criterion is not applicable.

\begin{table}[!htb]
\caption{The settings under which it is appropriate (\ok) or not (\notok) to evaluate the \textbf{Realism of Synthetic Data}, and any conditions associated with evaluation (\conditional).}
\label{tab:data-realism-guidelines}
\begin{tabular}{|p{0.13\linewidth}|p{0.15\linewidth}p{0.02\linewidth}p{0.21\linewidth}p{0.35\linewidth}|}
\hline
\multicolumn{1}{|l}{\textbf{Criteria}} & \multicolumn{1}{l}{\textbf{Category}} & \multicolumn{2}{c}{\textbf{Setting}} & \textbf{Notes} \\ \hline
\multirow{8}{\linewidth}{Realism of Synthetic Data} & \multirow{2}{\linewidth}{Evaluation Context} & \ok & General Purpose &  \\
 &  & \notok & Application-Based &  \\ \cline{2-5} 
 & \multirow{4}{\linewidth}{Data} & \ok & Tabular & \multirow{4}{\linewidth}{Further investigation needed to determine suitability for time series data.} \\
 &  & \conditional & Multivariate Time Series &  \\
 &  & \conditional & Univariate Time Series &  \\
 &  & \ok & Synthetic &  \\ \cline{2-5} 
 & Models & \ok & All Categories of Models &  \\ \cline{2-5} 
 & XAI Techniques & \conditional & All XAI Techniques & Technique must produce a synthetic neighbourhood as part of the explanation generation process \\ \hline
\end{tabular}
\end{table}

\section{XAI Complexity}\label{sec:complexity-slr}
This category contains two commonly applied criteria that measure XAI technique complexity and explanation complexity respectively.

\begin{figure}[!htb]
    \centering
    \includegraphics[scale=0.45]{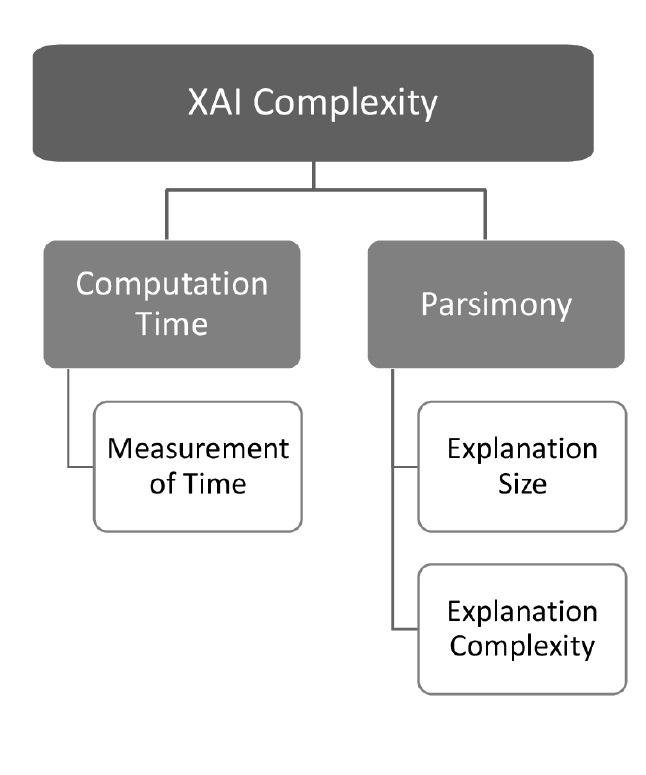}
    \caption{The three levels of the taxonomy elaborating the explanation quality dimension of \emph{XAI complexity}, the associated evaluation criteria, and the evaluation methods that may be used to test these explanation criteria. We identify at least one evaluation method for each criterion.}
    \label{fig:complexity-taxonomy}
\end{figure}

\subsection{Computation Time}
\vspace{0.5\baselineskip}

\noindent
\textbf{Evaluation Criterion.} This criterion measures the time required to generate an explanation. It is typically used as a stand-in for algorithm complexity.
This criterion is generally referred to as \textbf{runtime}~\cite{Lundberg2020} or  \textbf{computational complexity}~\cite{Maaroof2022}.
\vspace{0.5\baselineskip}

\noindent
\textbf{Evaluation Methods.} This criterion is simply the measurement of time required to generate an explanation. It may be measured as the time needed to compute a single explanation~\cite{Bjoerklund2022,Chen2022,Ferrettini2021,Hatwell2020,Maaroof2022,Mylonas2023}, the average time taken to generate explanations for a dataset~\cite{Buonocore2022,Doumard2023,Gatta2021,Mollas2022} or the total time required to compute explanations for a dataset~\cite{Veerappa2022,Lundberg2020}.
\vspace{0.5\baselineskip}

\noindent 
\textbf{Evaluation Guidelines.} {We analyse the evaluation of computation time in the literature, and find no restrictions associated with the evaluation (Table~\ref{tab:runtime-guidelines}).} 

\textit{Regarding Evaluation Context.} In literature, computation time has been assessed for both general-purpose and application-based settings. 

\textit{Regarding Data}. As the data is not directly used in the evaluation, it is not likely to limit evaluation. 

\textit{Regarding Model.} The model is not directly used in the evaluation, and will not limit evaluation. 

\textit{Regarding XAI Techniques.} Evaluation does not consider any aspect of the explanation itself. As long as start and end point for explanation generation can be defined, this criterion can be evaluated.

\begin{table}[h!!!]
\caption{The settings under which it is appropriate (\ok) or not (\notok) to evaluate \textbf{Computation Time}, and any conditions associated with evaluation (\conditional).}
\label{tab:runtime-guidelines}
\begin{tabular}{|p{0.13\linewidth}|p{0.15\linewidth}p{0.02\linewidth}p{0.21\linewidth}p{0.35\linewidth}|}
\hline
\multicolumn{1}{|l}{\textbf{Criteria}} & \multicolumn{1}{l}{\textbf{Category}} & \multicolumn{2}{c}{\textbf{Setting}} & \textbf{Notes} \\ \hline
\multirow{5}{\linewidth}{Computation Time} & \multirow{2}{\linewidth}{Evaluation Context} & \ok & General Purpose & \\
 &  & \ok & Application-Based &  \\ \cline{2-5} 
 & Data & \ok & All Forms of tabular Data & \\\cline{2-5} 
 & Models & \ok & All Categories of Models &  \\ \cline{2-5} 
 & XAI techniques & \ok & All XAI Techniques & \\ \hline
\end{tabular}
\end{table}

\subsection{Explanation Parsimony}\label{sec:parsimony}
\vspace{0.5\baselineskip}

\noindent
\textbf{Evaluation Criterion.}
Parsimony is a measure of the simplicity of an explanation, and is typically used as a proxy measure for explanation complexity and understandability.
%
In the literature, this criterion is known as \textbf{parsimony}~\cite{Cugny2022}, \textbf{conciseness}~\cite{Amparore2021,Doumard2023}, \textbf{sparsity}~\cite{Dai2022}, \textbf{compactness}~\cite{Cugny2022}, \textbf{interpretability}~\cite{Moradi2021}, \textbf{simplicity}~\cite{Maltbie2021} and \textbf{rule length}~\cite{Mollas2022,Mylonas2023}. We refer to this criterion as \textbf{``parsimony''}, as the other names imply association with size, complexity or human- and application-grounded evaluation.
\vspace{0.5\baselineskip}

\noindent
\textbf{Evaluation Methods.} There are two evaluation methods for this criterion:
\begin{itemize}
    \item \textit{Explanation Size}: This approach measures the size of explanations. For example, number of features with weights over a certain threshold~\cite{Amparore2021,Stevens2022,Dai2022}; number of features in a feature subset or decision rule~\cite{Cugny2022,Mollas2022,Moradi2021,Mylonas2023}; or number of separate items in the explanation, such as rulesets~\cite{Moradi2021} or prototype data points~\cite{Cugny2022}.
    \item \textit{Explanation Complexity}: This approach measures the complexity of explanations. For example, the average distance between explanations for a set of prototype points~\cite{Cugny2022}, overlap in explanation coverage~\cite{Moradi2021} and distribution~\cite{Doumard2023} or entropy~\cite{Maltbie2021} of feature importance.
\end{itemize}

\noindent 
\textbf{Evaluation Guidelines.} {We analyse the relevant context of existing evaluation methods (referred to as evaluation context), data, (predictive) model and XAI techniques, and then derive evaluation guidelines for assessing explanation parsimony. These are summarised in Table~\ref{tab:parsimony-guidelines}.} 

\textit{Regarding Evaluation Context.} In the surveyed literature, parsimony is evaluated in both general-purpose and application-based contexts. 

\textit{Regarding Data.} Parsimony is sometimes measured as explanation size, i.e. the number of features deemed important in the explanation. It is unclear what would be considered parsimonious when using univariate data, and requires further investigation.

\textit{Regarding Model.} The model is not directly used in evaluation, and choice of model does not limit evaluation.

\textit{Regarding XAI Techniques.} Parsimony is hard to define for feature profiling explanations, such as Partial Dependency Plots. It is unclear what would be considered parsimonious in such an explanation.


\begin{table}[!htb]
\caption{The settings under which it is appropriate (\ok) or not (\notok) to evaluate \textbf{Explanation Parsimony}, and any conditions associated with evaluation (\conditional).}
\label{tab:parsimony-guidelines}
\begin{tabular}{|p{0.13\linewidth}|p{0.15\linewidth}p{0.02\linewidth}p{0.21\linewidth}p{0.35\linewidth}|}
\hline
\multicolumn{1}{|l}{\textbf{Criteria}} & \multicolumn{1}{l}{\textbf{Category}} & \multicolumn{2}{c}{\textbf{Setting}} & \textbf{Notes} \\ \hline
\multirow{12}{\linewidth}{Explanation Parsimony} & \multirow{2}{\linewidth}{Evaluation Context} & \ok & General Purpose &  \\
 &  & \ok & Application-Based &  \\ \cline{2-5} 
 & \multirow{4}{\linewidth}{Data} & \ok & Tabular & \multirow{4}{\linewidth}{Current evaluation methods focus on the distinct variables in the data -- unclear how this method would translate for univariate data.} \\
 &  & \ok & Multivariate Time Series &  \\
 &  & \conditional & Univariate Time Series &  \\
 &  & \ok & Synthetic &  \\ \cline{2-5} 
 & Models & \ok & All Categories of Models &  \\ \cline{2-5} 
 & \multirow{5}{\linewidth}{XAI techniques} & \ok & Feature Attribution & \multirow{5}{\linewidth}{Further investigation needed to determine applicability for feature profiling explanations. Appropriate metrics may vary across explanation types.} \\
 &  & \multirow{2}{\linewidth}{\conditional} & \multirow{2}{\linewidth}{Feature Profiling} &  \\
 &  &  &  &  \\
 &  & \ok & Feature Sets &  \\
 &  & \ok & Rule-Based &  \\ \hline
\end{tabular}
\end{table}

\section{Explanation Safety}\label{sec:fairness-slr}
We identified two criteria that were related to the explanation safety. These criteria measured whether an explanation could be applied without risk of harm.

\begin{figure}[!htb]
    \centering
    \includegraphics[scale=0.45]{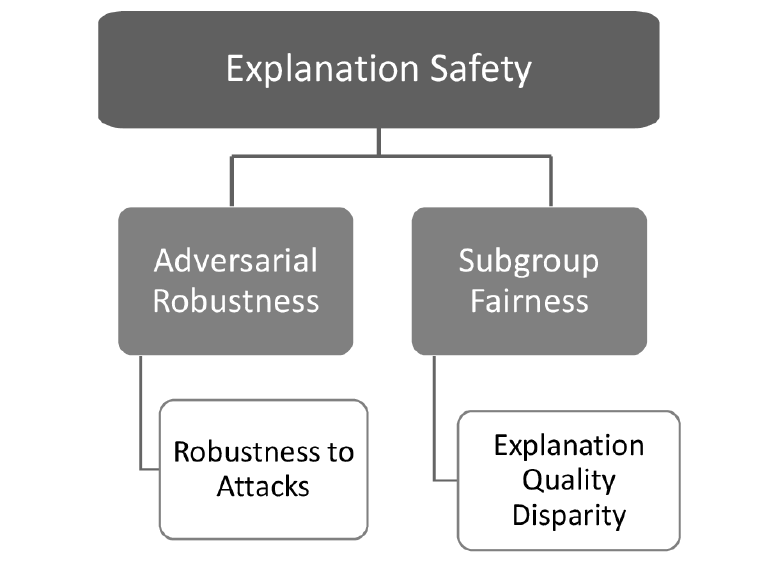}
    \caption{The three levels of the taxonomy elaborating the explanation quality dimension of \emph{explanation safety}, the associated evaluation criteria, and the evaluation methods that may be used to test these explanation criteria. We identify at least one evaluation method for each criterion.}
    \label{fig:safety-taxonomy}
\end{figure}

\subsection{Adversarial Robustness}~\label{sec:adversarial-robustness-def}
\vspace{0.5\baselineskip}

\noindent
\textbf{Evaluation Criterion.} This criterion evaluates how well an XAI technique is able to identify sensitive variables used by the model when explaining data points formulated to hide the influence of sensitive variables.
\vspace{0.5\baselineskip}

\noindent
\textbf{Evaluation Methods.} Adversarial data points are generated, and the most important feature according to the explanation is recorded. The percentage of adversarial data points for which the explanation is deceived is used to measure robustness. 
\vspace{0.5\baselineskip}

\noindent
\textbf{Evaluation Guidelines.} We cannot extract guidelines for assessing adversarial robustness. This criterion is evaluated in only one paper~\cite{Vres2022}. 
The experimental settings used in the surveyed paper are limited, and the other criterion in this dimension is not similar enough to draw shared conclusions. Evaluating the adversarial robustness of explanations is a research gap that future works should explore.
\subsection{Subgroup Fairness}~\label{sec:subgroup-fairness-def}

\noindent
\textbf{Evaluation Criterion.} This criterion measures how fair explanations are across demographic subgroups in the data, where fairness relates to the quality of explanations.
This criterion is most commonly called \textbf{fairness}~\cite{Balagopalan2022,Dai2022}.
Since ``fairness'' is often used to refer to bias in machine learning models and is an overarching concept in the field of AI and ML, we refer to this criterion more specifically as \textbf{``subgroup fairness''}.
\vspace{0.5\baselineskip}

\noindent
\textbf{Evaluation Methods.} Subgroup fairness is measured as the difference in explanation quality for different subgroups in the data~\cite{Balagopalan2022,Dai2022}. For example, the assessor may calculate the average explanation parsimony for each race of people present in the data. Then, the differences in explanation parsimony between these subgroups is measured as fairness~\cite{Balagopalan2022,Dai2022}. 
\vspace{0.5\baselineskip}

\noindent
\textbf{Evaluation Guidelines.} We cannot extract guidelines for assessing subgroup fairness. This criterion relies on other evaluation criteria and so it is impossible to derive guidelines for subgroup fairness alone. 
Thus, restrictions on evaluating subgroup fairness depend on the criterion being used to determine fairness, and we cannot definitively provide guidelines for subgroup fairness evaluation.

\section{Application Utility}\label{sec:utility-slr}
We identified three criteria that related to determining the usefulness of explanations in a real-world setting.

\begin{figure}[!htb]
    \centering
    \includegraphics[scale=0.45]{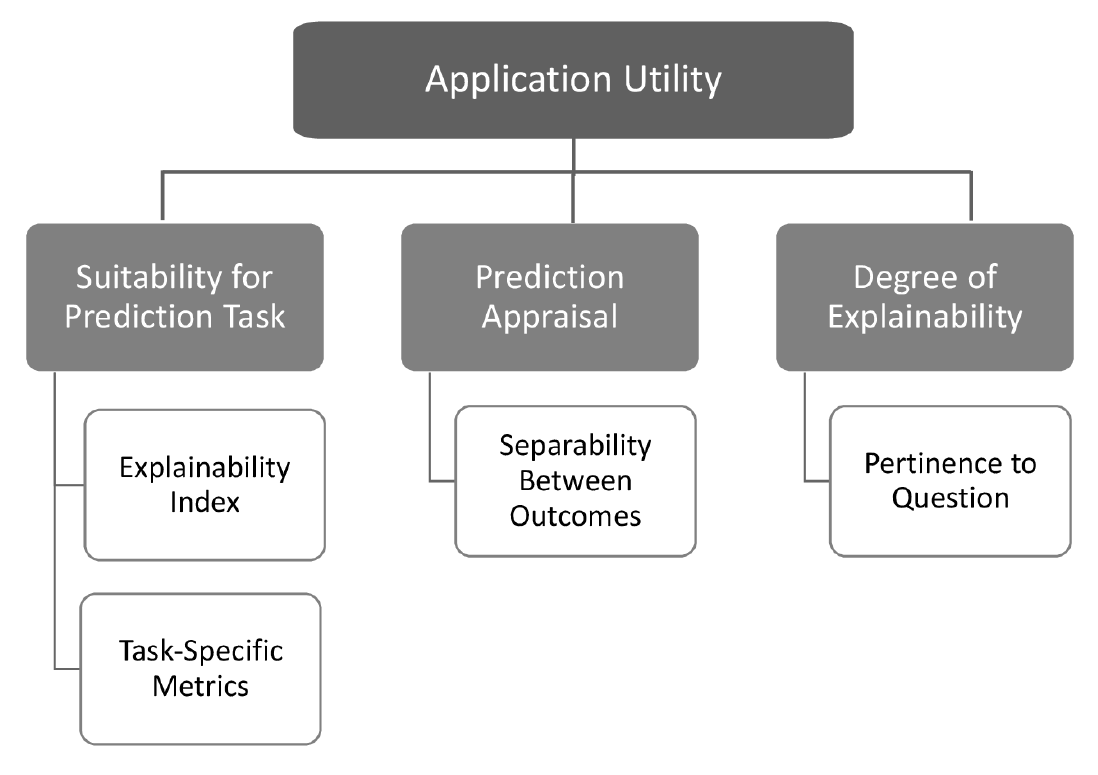}
    \caption{The three levels of the taxonomy elaborating the explanation quality dimension of \emph{application utility}, the associated evaluation criteria, and the evaluation methods that may be used to test these explanation criteria. We identify at least one evaluation method for each criterion.}
    \label{fig:utility-taxonomy}
\end{figure}

\subsection{Suitability for Prediction Task}\label{sec:task-suitability}
\vspace{0.5\baselineskip}

\noindent
\textbf{Evaluation Criterion.}
This criterion measures how well an XAI technique performs at some task-specific objective.
\vspace{0.5\baselineskip}

\noindent
\textbf{Evaluation Methods.}
The name, definition and measurement of this criterion are heavily dependent on the domain context. We identify one evaluation method that is transferable across domains, and three that are not. Thus, we cannot categorise together common evaluation methods and instead present each form of this criterion separately.

\textit{Explainability Index.} Melo et al~\cite{Melo2022} apply an \textbf{``explainability index''} to measure the usefulness of XAI techniques in an educational context. This is simply a count of how many task-based requirements are met by explanations produced by different techniques. Requirements identified by Melo et al~\cite{Melo2022} include, ``Shows interaction between features,'' and, ``Provides local explanations.'' This ``checklist'' approach of evaluation is transferable across domains, the requirements are specific to the task and domain.

\textit{Task-Specific Metrics.} In their work on remaining useful life (RUL) prediction for jet engines, Baptista, Goebel and Henriques~\cite{Baptista2022} evaluate explanations across three domain-specific criteria, using domain-specific metrics:
\begin{itemize}
    \item Monotonicity: how well an explanation captures the increasing  or decreasing trend of a predictor. 
    \item Trendability: measures the extent to which an explanation captures the same shape across a group of units.
    \item Prognosability: measures the variance of the explanation at the end of life for a set of units. 
\end{itemize}

This evaluation method is highly specific to the domain, so this list of methods is not comprehensive.
\vspace{0.5\baselineskip}

\noindent
\textbf{Evaluation Guidelines.} We cannot extract guidelines for assessing this criterion, as it is highly specific to the domain, and may take many forms. 
\subsection{Prediction Appraisal}
\vspace{0.5\baselineskip}

\noindent
\textbf{Evaluation Criterion.} This criterion evaluates how easy it is to identify true and false predictions (i.e. model correctness) given an explanation.

\vspace{0.5\baselineskip}

\noindent
\textbf{Evaluation Methods.} Arslan et al~\cite{Arslan2022} use cluster feature attribution explanations, then extract decision rules from these clusters. This criterion is evaluated as how well the clusters and decision rules discriminate between true and false predictions. 
This method does not take the target variable as the ground truth, as a predictive model would, and they do not take the set of predictions for the dataset, as a surrogate model would. Instead, the target ``variable'' is whether the model's prediction is correct, i.e. whether the prediction matches the true value. 
\vspace{0.5\baselineskip}

\noindent
\textbf{Evaluation Guidelines.} We cannot extract guidelines for assessing prediction appraisal. This criterion is only evaluated in one work~\cite{Arslan2022} and is evaluated with limited experimental settings. 
While we can attempt to draw conclusions based on the evaluation method and settings that are used by Arslan et al~\cite{Arslan2022}, we believe that work on this criterion is not mature enough to draw firm guidelines. No other criteria in this dimension are similar enough for us to draw shared conclusions. 
\subsection{Degree of Explainability}\label{sec:dox}
\vspace{0.5\baselineskip}

\noindent
\textbf{Evaluation Criterion.} This criterion measures the informativeness of an explanation. While explanation usefulness is typically evaluated through human- and application-grounded evaluation, this criterion is functionally-grounded.
\vspace{0.5\baselineskip}


\noindent
\textbf{Evaluation Methods.} Sovrano and Vitali~\cite{Sovrano2022} measure the degree of explainability (DoX) as an explanation's ability produce new understandings based on archetypal questions. The authors predefine a set of archetypal questions and use support material relevant to the subject to identify the details relevant to the domain. Pre-trained language models are used to determine the pertinence of the explanation~\cite{Sovrano2022}.
\vspace{0.5\baselineskip}

\noindent
\textbf{Evaluation Guidelines.} We cannot extract guidelines for assessing DoX. DoX is proposed and evaluated in only one work~\cite{Sovrano2022} and is evaluated under limited settings. While we can draw conclusions based on the evaluation methods and settings that are used by Sovrano and Vitali~\cite{Sovrano2022}, we believe that work on this criterion is not mature enough to draw firm guidelines, and no other criterion in this dimension is similar enough to draw shared conclusions. Furthermore, the method of evaluating DoX is elaborate and it is unclear how design decisions regarding each component of the evaluation may impact evaluation. 








\section{Discussion}\label{sec:discussion}
In this paper, we investigated how post hoc XAI techniques can be evaluated in settings that use tabular data. We developed a taxonomy of evaluation criteria and evaluation methods, then derived guidelines on how and when each criterion should be evaluated. 

\paragraph{Application of findings.} We apply our findings to the example scenarios in Section~\mbox{\ref{sec:intro}}. 
In the first scenario, University A is, at this stage, attempting to narrow down its options to identify the most appropriate XAI techniques for its aims---i.e., the university is attempting to determine the \textit{suitability of XAI techniques for the prediction task} at hand (Section~\ref{sec:task-suitability}). Given that University A has not yet determined concrete metrics to evaluate explanations against, it should choose the \textit{explainability index method} of evaluation (Section~\ref{sec:task-suitability}). That is, it should develop a ``checklist'' of atomic requirements that the desired solution should provide and score XAI techniques against this index. It can thus develop a shortlist of techniques for use, and subsequent evaluation of these techniques against other criteria will allow the university to narrow down its options further.

In the second scenario, Company B aims to evaluate the faithfulness of local, post hoc explanations for PPA models. More specifically, it intends to determine 1) how complete an explanation is with respect to the features that produce a given prediction (i.e. are other features needed to produce the same prediction) and 2) how necessary a given feature or a set of features is to a prediction (i.e. can the same prediction be made without a given feature or set of features). In our taxonomy, 
the two relevant evaluation criteria are \textit{Sufficiency} (Section~\ref{sec:sufficiency}) and \textit{Necessity} (Section~\ref{sec:necessity}), respectively. By referring to the guidelines for these two criteria (Table~\mbox{\ref{tab:necessity-guidelines}} and Table~\mbox{\ref{tab:sufficiency-guidelines}}), it becomes apparent that there are no procedures for evaluating these criteria when using time series data. Although business process event logs are not strictly time series data, they share many characteristics, and the restrictions applicable to time series data may also apply here. Therefore, Company B may instead choose to evaluate a similar criterion, such as internal correctness (Section~\ref{sec:internal-corr}).

\paragraph{Research gaps to address in future work.}


{Firstly, we did not include evaluation metrics in the taxonomy, only high-level evaluation methods. In the surveyed literature, while evaluation methods are typically consistent, metrics vary greatly. For example, one paper may use the RMSE metric and another may use the precision metric for the same evaluation method. Thus, we exclude evaluation metrics from our work but suggest that guidelines on determining appropriate metrics may be an avenue for future work.}

{Secondly, we identified several criteria in the taxonomy that require further study. In particular, we have identified five, relatively new evaluation criteria which require further investigation. These criteria are concerned with the application utility and explanation safety dimensions. The usefulness of evaluating these criteria, as well as the practicality of their application, are still very much unknown. Investigating these criteria is a key direction for future research.} {We identify several other criteria that require further investigation:}
\begin{itemize}
    \item {Completeness: Further research is needed to determine how completeness should be evaluated when using tabular data.}
    \item {Necessity: Further research is needed to investigate the practicality of evaluating necessity in real-world settings and the appropriate methods of evaluation when using time series data.}
    \item {Sufficiency: Further research is needed to investigate the practicality of evaluating necessity in real-world settings and the appropriate methods of evaluation when using time series data.}
    \item {Descriptive Accuracy: Further research is needed to investigate the practicality of evaluating necessity in real-world settings, the appropriate methods of evaluation when data has continuous features and the appropriate methods of evaluation when using time series data.}
    \item {Sensitivity: Further research is needed to investigate the ``clustering'' method of evaluation, which is very similar to a method that evaluates external correctness.}
    \item {Synonymity: Further research is needed to investigate the practicality of evaluation in application-based settings.}
    \item {Explanation Coverage: Further research is needed to investigate usefulness in application-based settings, the appropriate methods of evaluation when using time series data and the appropriate methods of evaluation for XAI techniques that are not rule-based.}
    \item {Explanation Realism: Further research is needed to investigate the applicability of this criterion for feature profiling and feature set explanations.}
    \item {Realism of Synthetic Data: Further research is needed to investigate suitability for time series data.}
    \item {Explanation Parsimony: Further research is needed to investigate the appropriate methods of evaluation when using univariate data and the applicability of explanation parsimony for feature profiling explanations.}
\end{itemize} 

{We have also identified gaps or inconsistencies in research evaluating well-established and commonly evaluated criteria. In particular, we find that internal correctness evaluation in real-world settings may not be possible. This is particularly notable as internal correctness is highly dependent on all components of the technical pipeline~\mbox{\cite{Gaudel2022,Mollas2022a,Naretto2022,Schlegel2021,Stevens2022,Velmurugan2023}}, and evaluation in full context is needed to determine explanation correctness in a real-world setting. Future work should address this. Moreover, explanations provided by different XAI techniques often have distinct meanings. For example, LIME explanations indicate feature importance, while SHAP explanations indicate contribution. We suggest that future work should take these into account while evaluating the internal correctness of explanations.} 

{Finally, future work must also address limitations in our work. In this paper, we have chosen to develop guidelines for a relatively limited range of settings, to better enable us to analyse these settings in depth. Therefore, we have made several notable exclusions when selecting papers to review.} For example, we excluded evaluations of counterfactual algorithms, which have gained increasing attention. 
However, counterfactual explanations have specific criteria that they must meet~\cite{Chou2021fusion}. 
{Furthermore, we have excluded global explanations and inherently interpretable, ``transparent'' models.}  
{Our work also focuses exclusively on tabular data. We suggest that our guidelines should be adapted and extended for other types of data by future work.}

\section{Conclusion}\label{sec:conclusion}
In this paper, we have investigated the functionally-grounded evaluation of post hoc XAI techniques in settings that use tabular data. While existing surveys and taxonomies generally do not differentiate between different types of data when considering XAI evaluation, tabular data contain specific characteristics that affect evaluation and cannot be overlooked. 
We have performed in-depth analysis to investigate the evaluation of XAI techniques in settings that use tabular data, and have derived guidelines to support future evaluations of XAI techniques. Our work has made the following contributions:
\begin{itemize}
    \item identification of functionally-grounded criteria that explanations must meet;
    \item identification of evaluation methods and metrics that have been used to assess explanations against the identified criteria; and
    \item identifcation of evaluation methods and their appropriateness for a technical context for each criterion.
\end{itemize}
{Additionally, we have also identified several key gaps and areas of improvement in XAI evaluations that future works should address. 
This work contributes to the body of knowledge on explainable AI evaluation. Through our in-depth examination of XAI protocols, we have laid the groundwork for future research on XAI evaluation.}





\printcredits

\section*{Competing Interests}
The authors declare that they have no known competing financial interests or personal relationships that could have appeared to influence the work reported in this paper.

\section*{Funding}
The reported research is part of a Ph.D. project supported by the Australian Research Training Program, which has funded the first author's scholarship. The fifth author is supported by a Science and Engineering Faculty scholarship and a Centre for Data Science top up scholarship at Queensland University of Technology (QUT), Australia.

\bibliographystyle{cas-model2-names}

\bibliography{cas-refs}

\bio{}
\textbf{Mythreyi Velmurugan} received her PhD from the Queensland University of Technology in 2024. Her research interests include the use of artificial intelligence and explainable artificial intelligence in the field of Business Process Management, as well as the evaluation and appropriate use of explainable AI technologies.
\endbio

\bio{}
\textbf{Chun Ouyang} received her PhD from the University of South Australia in 2004. She is currently an associate professor at the Queensland University of Technology, and a co-founder and leader of the `Explainable Analytics for Machine Intelligence' (XAMI) research initiative. A key focus of her research interests is on process mining, especially data-driven process analytics using data mining and explainable machine learning techniques.


\endbio

\bio{}
\textbf{Yue Xu} received her PhD from the University of New England in 2000. She is currently an Associate Professor at the Queensland University of Technology. She has been an active researcher in the areas of data analytics, machine learning and knowledge discovery for more than 20 years.
\endbio

\bio{}
\textbf{Renuka Sindhgatta} received her PhD from the University of Wollongong. She is currently a Senior Technical Staff Member and Manager at IBM Research India. She works across IBM's Research and product teams, and focuses on supporting automation via a hybrid workforce of intelligent agents and humans to accelerate the digital transformation of organisations.
\endbio

\bio{}
\textbf{Bemali Wickramanayake} is currently a PhD candidate at the Queensland University of Technology. Her research interests include the use of artificial intelligence and explainable artificial intelligence in the field of Business Process Management, particularly in explaining deep learning-based process prediction models intrinsically.
\endbio

\bio{}
\textbf{Catarina Moreira} received her PhD from the University of Lisbon. She is currently a Lead Research Scientist at the University of Technology Sydney and an Adjunct Professor at the Queensland University of Technology. She is also the UNESCO Co-Chair on Artificial Intelligence and Extended Reality, as well as a co-founder and leader of the `Explainable Analytics for Machine Intelligence' (XAMI) research initiative. 
\endbio

\end{document}